\documentclass[aps,prl,twocolumn,showpacs,superscriptaddress]{revtex4-2}

\usepackage{amsmath}
\usepackage{graphicx}
\usepackage{lmodern}
\usepackage{amsmath}

\usepackage{color}
\usepackage{amssymb}
\usepackage{bm}
\usepackage{braket}
\usepackage{mathtools}
\usepackage{slashed}

\newcommand{\tr}{\text{Tr}}

\newcommand{\ke}{Kekul\'{e} }

\usepackage[dvipsnames]{xcolor}
\usepackage[colorlinks=true]{hyperref}
\hypersetup{
    colorlinks=true,
    linkcolor=blue,
    citecolor=blue,
    urlcolor=blue
}

\makeatletter
\def\maketitle{
\@author@finish
\title@column\titleblock@produce
\suppressfloats[t]}
\makeatother
\begin{document}

\title {Emergent spacetime supersymmetry at 2D fractionalized quantum criticality}
\date{\today}
\author{Zhengzhi Wu}
\thanks{These two authors contributed equally to this work.}
\affiliation{Institute for Advanced Study, Tsinghua University, Beijing 100084, China}
\affiliation{Rudolf Peierls Centre for Theoretical Physics, Parks Road, Oxford, OX1 3PU, UK}
\author{Zhou-Quan Wan}
\thanks{These two authors contributed equally to this work.}
\affiliation{Center for Computational Quantum Physics, Flatiron Institute, New York, NY 10010, USA}
\author{Shao-Kai Jian}
\email{sjian@tulane.edu}
\affiliation{Department of Physics and Engineering Physics, Tulane University, New Orleans, Louisiana 70118, USA}
\author{Hong Yao}
\email{yaohong@tsinghua.edu.cn}
\affiliation{Institute for Advanced Study, Tsinghua University, Beijing 100084, China}

\begin{abstract}
While experimental evidence for spacetime supersymmetry (SUSY) in particle physics remains elusive, condensed matter systems offer a promising arena for its emergence at quantum critical points (QCPs). 
Although there have been 
a variety of proposals for emergent SUSY at symmetry-breaking QCPs, the emergence of SUSY at fractionalized QCPs remains largely unexplored. 
Here, we demonstrate emergent space-time SUSY at a fractionalized QCP in the Kitaev honeycomb model with Su-Schrieffer-Heeger (SSH) spin-phonon coupling. Specifically, through numerical computations and analytical analysis, we show that the anisotropic SSH-Kitaev model hosts a fractionalized QCP between a Dirac spin liquid and an incommensurate/commensurate valence-bond-solid phase coexisting with $\mathbb{Z}_2$ topological order. 
A low-energy field theory incorporating phonon quantum fluctuations reveals that this fractionalized QCP features an emergent $\mathcal{N}=2$ spacetime SUSY. 
We further discuss their universal experimental signatures in thermal transport and viscosity, highlighting the concrete lattice realization of emergent SUSY at a fractionalized QCP in 2D. 
\end{abstract}
\maketitle

{\bf Introduction:} 
Supersymmetry (SUSY), a fundamental spacetime symmetry relating bosons and fermions \cite{GERVAIS1971632,WESS197439,DIMOPOULOS1981150}, has been extensively explored in high-energy physics, but has not been observed in nature at accessible energy scales. 
Condensed matter systems, however, provide a promising alternative platform for exploring SUSY \cite{PhysRevLett.126.206801,PhysRevLett.90.120402,PaulFendley_2003,PhysRevLett.117.166802,PhysRevLett.126.236802,XiaoYang_2004,PhysRevLett.95.046403,PhysRevLett.101.146406,PhysRevB.84.115124,PhysRevLett.126.236802,PhysRevB.100.195146,PhysRevB.103.085130,Cai2022,PhysRevB.110.165124,6722-tf9c,PhysRevResearch.6.043273}. 
Crucially, spacetime SUSY can emerge dynamically at quantum critical points between symmetric and symmetry-broken phases~\cite{PhysRevLett.52.1575,PhysRevB.76.075103,doi:10.1126/science.1248253,Ponte_2014,PhysRevLett.114.237001,PhysRevLett.114.090404,PhysRevLett.115.166401,PhysRevLett.118.166802,PhysRevLett.119.107202,PhysRevB.103.014435,PhysRevLett.133.223401}, providing a natural arena for studying its consequences by tuning microscopic parameters.
Despite this progress, SUSY in fascinating fractionalized settings, which typically involve topological order with deconfined gauge fields and fractionalized particles \cite{ANDERSON1973153,PhysRevB.35.8865,PhysRevLett.96.110404,PhysRevLett.96.110405}, remains largely unexplored.

More specifically, achieving emergent SUSY at fractionalized quantum critical points (QCPs) \cite{doi:10.1126/science.1212207,PhysRevB.94.085134,PhysRevLett.117.210401,doi:10.1073/pnas.1806338115,Vojta_2018,PhysRevLett.125.257202} has been an open problem and poses significant theoretical challenges, owing to the intrinsic difficulties of reliably treating deconfined gauge fields and frustrated spin interactions at phase transitions between quantum spin liquids (QSLs) and symmetry-breaking phases. 
However, the Kitaev honeycomb model \cite{KITAEV20062} (and other related models \cite{PhysRevD.68.065003,hongchiral,PhysRevLett.99.196805,Baskaran2008SpinS,PhysRevLett.102.217202,PhysRevB.79.024426,Wu2009GammaMatrix,PhysRevB.83.180412,PhysRevLett.107.087205,PhysRevLett.114.157202,Chulliparambil2020SixteenfoldMicroscopic,Jin2023SixteenfoldExactlySolvable,PhysRevLett.130.156701,PhysRevLett.133.236504,Eck2025BraidedFusionGeneralizations}) offers a promising platform to overcome these challenges. 
As a paradigmatic solvable model hosting $\mathbb{Z}_2$ QSLs with potential material realizations \cite{PhysRevLett.102.017205,annurev:/content/journals/10.1146/annurev-conmatphys-031115-011319,annurev:/content/journals/10.1146/annurev-conmatphys-033117-053934,wen2019experimental,TREBST20221}, the Kitaev model has conserved deconfined $\mathbb{Z}_2$ gauge fields. 
While additional interactions (e.g., Heisenberg couplings) typically endow these gauge fields with dynamics, we find that certain spin-phonon couplings can preserve the $\mathbb{Z}_2$ gauge structure while simultaneously driving a fractionalized QCP with emergent SUSY via a spin-Peierls instability \cite{PhysRevB.10.4637,Kuboki_1987,PhysRevLett.70.3651,PhysRevB.57.R14004,PhysRevLett.83.195,PhysRevLett.62.1694,PhysRevB.42.4568,PhysRevLett.89.037204,PhysRevB.60.6566,PhysRevLett.115.177205,PhysRevB.104.165133,PhysRevB.104.035126,Seifert2024,PhysRevB.110.125130,ferrari2024stabilityalgebraicspinliquids,2024nonequilibrium}. 
Crucially, the QCPs of this instability remain far less understood in high dimensions than the resulting ordered phases. 
Moreover, spin-phonon coupling in the Kitaev model has attracted significant interest beyond the spin-Peierls instability \cite{PhysRevLett.121.147201,PhysRevX.8.031032,PhysRevResearch.2.033180,PhysRevB.101.035103,PhysRevB.106.024413,dantas2024phonondynamicssitedisorderedkitaev,ljlq-1kgl}, as it is believed to underlie several experimental observations in Kitaev materials, such as the thermal Hall conductivity \cite{Kasahara2018,Chen2024}.  

\begin{figure*}[t]
  \includegraphics[width=1.0\textwidth]{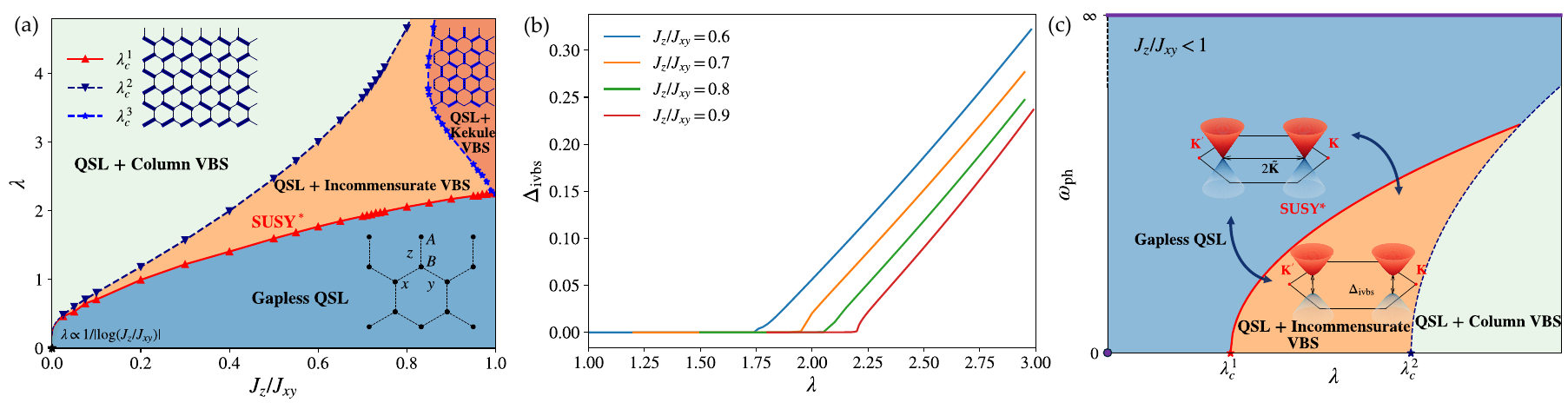}
 \caption{(a) The quantum phase diagram in the adiabatic limit. The horizontal and vertical axes represent the anisotropy parameter $J_z/J_{xy}$ and the dimensionless spin-phonon coupling $\lambda$, respectively. For weak coupling $\lambda < \lambda_c^{1}$ (red line), the ground state is a gapless spin liquid with a single Dirac cone with both positive- and negative-energy branches. As $\lambda$ increases, the system first enters an incommensurate VBS phase with topological order, and eventually transitions into a columnar VBS phase with period two when $\lambda > \lambda_c^{2}$ (dark blue dashed line). In the incommensurate regime, there inevitably exist regions of commensurate VBS phases whose phase boundaries are difficult to resolve numerically when the periodicity is large. Here we explicitly show the case of period three, corresponding to a Kekulé VBS order, with $\lambda_c^{3}$ marking its phase boundary. The numerical calculations are performed on a finite lattice with $2\times 240 \times 120$ sites, and phase boundaries $\lambda_c^{2,3}$ are identified by comparing the energies of competing phases. $\lambda_c^1$ are identified by the linear extrapolation of the order parameter $\Delta_\text{ivbs}$ \cite{SupMat}. 
 (b) The incommensurate VBS order parameter $\Delta_\text{ivbs}\equiv \big|\tfrac{1}{N}\sum_{i\in A} e^{-2i  \tilde {\mathbf{K}} \cdot\mathbf r_i} X_{\langle ij\rangle\in z}\big|$ near the critical point $\lambda_c^1$, which clearly exhibits a continuous phase transition.  Here $\tilde{\mathbf{K}}$ denotes the momentum of the Dirac cone, which is illustrated as two red positive-energy branches in the full Brillouin zone (BZ) in Fig. 1(c).
 The linear onset indicates the presence of $|\Delta_\text{ivbs}|^3$ terms in the free energy, originating from the Dirac cone.
 (c) Schematic phase diagram at a finite phonon frequency $\omega_{\text{ph}}$ which is supported by 
 perturbative calculations \cite{SupMat}. The fractionalized quantum critical line between the gapless QSL and the incommensurate VBS QSL belongs to the $\text{SUSY}^*$ universality class. In the limit $\omega_{\text{ph}} \to \infty$, the spin degrees of freedom are decoupled from phonons.
 }
 \label{AdiaPhaseDiag}
\end{figure*}

In this Letter, we propose that the Kitaev honeycomb model with spin-phonon coupling, which modulates the strength of Kitaev exchange interactions in proportion to phonon displacements, can host emergent SUSY at fractionalized quantum criticality. 
Combining Lieb's theorems and large-scale numerical computations, we first obtain the phase diagram in the adiabatic limit, revealing transitions from a Dirac QSL to incommensurate or commensurate VBS order coexisting with $\mathbb{Z}_2$ topological order. 
Significantly, incorporating phonon quantum fluctuations away from the adiabatic limit, we construct a low-energy theory of Dirac fermions coupled to the VBS order parameter and show that these transitions flow to an emergent
$\mathcal{N}=2$ spacetime $\text{SUSY}^*$ fixed point. 
This provides a concrete lattice realization of SUSY at a fractionalized QCP, with universal signatures in thermal transport and viscosity. 
Our proposal of $\text{SUSY}^*$ fixed points is intrinsically different from previous ones \cite{doi:10.1126/science.1248253,doi:10.1126/sciadv.aau1463}, since our model neither resides at the boundary of a topological phase nor requires nonlocal interactions. 
Furthermore, unlike Ref.~\cite{doi:10.1126/science.1248253}, we do not need to impose particle-hole symmetry of the order-parameter fluctuations to avoid a multicritical point, since it is automatically enforced by an inversion symmetry in our model.

{\bf Model:} We consider the following SSH-Kitaev model on the honeycomb lattice:
\begin{equation}
    \hat{H}=\sum_{\langle ij\rangle\in \mu}(J_{\mu}+g\hat{X}_{\langle ij\rangle}) \tau_i^{\mu}\tau_j^{\mu}+\sum_{\langle ij\rangle}\frac{\hat{P}^2_{\langle ij\rangle}}{2m}+\frac{k}{2}\hat{X}^2_{\langle ij\rangle},
\end{equation}
where each bond $\langle ij\rangle$ is labeled with $\mu=x,y,z$ according to its direction, as illustrated in the inset of Fig.~\ref{AdiaPhaseDiag}(a). $\hat{X}_{\langle ij\rangle}$ is the phonon field on the nearest-neighbor bond ${\langle ij\rangle}$ and $\hat{P}_{\langle ij\rangle}$ is the conjugate phonon momentum. Here we consider the simplest Einstein phonon with phonon frequency $\omega_{\text{ph}}=\sqrt{\frac{k}{m}}$.   Using the Majorana fermion representation $\tau^{\mu}_{i}=i\hat{c}_i^{\mu}\hat{c}_i$ \cite{KITAEV20062}, where $\hat{c}_i^{\mu},\hat{c}_i$ are Majorana fermions, the Hamiltonian can be rewritten as
\begin{equation}
    \hat{H}_f=\sum_{\langle ij\rangle\in \mu}\left[\hat{u}_{\langle ij\rangle}(J_{\mu}+g\hat{X}_{\langle ij\rangle})\right]\left(i\hat{c}_i\hat{c}_j\right)+\sum_{\langle ij\rangle}\frac{\hat{P}^2_{\langle ij\rangle}}{2m}+\frac{k}{2}\hat{X}^2_{\langle ij\rangle}.
\end{equation}
The $\mathbb{Z}_2$ gauge fields $\hat{u}_{\langle ij\rangle\in\mu}=i\hat{c}^{\mu}_i\hat{c}^{\mu}_j$ are conserved, implying that the model's dynamics is governed solely by itinerant Majorana fermions and phonon fields.  
To realize a SUSY critical point, the model should exhibit a continuous phase transition with an equal number of Dirac cones and complex bosons, which is fulfilled here as we demonstrate below.

To gain intuition, we start from the 
$C_3$ symmetric point $J_x\!=\!J_y\!=\!J_z$. 
In the adiabatic limit $m\rightarrow\infty$, the phonon field $\hat{X}_{\langle ij\rangle}$ becomes a classical field, and there is a phase transition between the gapless quantum spin liquid ($g$=0) and the $\mathbb{Z}_2$ topological order ($g$$\rightarrow$$\infty$) with the increase of $g$. 
The $\mathbb{Z}_2$ topological order phase also exhibits a valence-bond-solid long-range order with a \ke pattern, which breaks the $C_3$ rotation symmetry and gaps out the Dirac cone located at the corner $\bf{K}$ of the Brillouin zone. 
This phase transition can be deduced to be continuous in the adiabatic limit from a Landau-Ginzburg free energy analysis, where the cubic term of the order parameter $X^3_\mathbf{K}+\text{h.c.}$ (the order parameter $X_\mathbf{K}$ is the Fourier component of $X_{\langle ij\rangle}$ at momentum $\mathbf{K}$) is overcome by the non-analytic term $|X_\mathbf{K}|^3$. 
The non-analytic term $|X_\mathbf{K}|^3$ arises from integrating out gapless fermions $\hat{c}_i$, analogous to the mechanism of the previously investigated fermion-induced QCP \cite{Li2017,
PhysRevB.96.195162,PhysRevB.101.085105}.  
However, with the inclusion of quantum fluctuations, or equivalently a large but finite $m$, the cubic term becomes relevant at a presumably continuous phase transition \cite{Li2017}, which means the phase transition becomes first-order with an infinitesimal deviation from the $m\rightarrow\infty$ limit.

Building on insights from the isotropic limit, we focus on the anisotropic regime to eliminate the cubic term $X^3_{\mathbf{K}}+\text{h.c.}$ and thus access a continuous phase transition beyond the adiabatic limit. We choose the coupling constants as: $J_x=J_y=J>0$ and $J_z=aJ$ with $0<a<1$, which produces a gapless spin liquid when $g=0$. 
We first analyze its phase diagram in the adiabatic limit, which paves the way for further controlled analysis of quantum fluctuations. 
After a rescaling of the phonon field: $X_{\braket{ij}}\rightarrow X_{\braket{ij}}J/g$, $\hat{H}_f$ (with $J=1$) can be simplified to a form with only one tuning parameter $\lambda=\frac{g^2}{k}$: 
\begin{eqnarray}
    \hat{H}_f&=\sum_{\langle ij\rangle\in \mu}\hat{u}_{\langle ij\rangle}(a_{\mu}+X_{\langle ij\rangle})\left(i\hat{c}_i\hat{c}_j\right)+\sum_{\langle ij\rangle}\frac{X^2_{\langle ij\rangle}}{2\lambda},~
    \label{adiabatic}
\end{eqnarray}
where $a_{x}=a_y=1$ and $a_z=a$ represent the anisotropy of the Kitaev couplings.

{\bf The quantum phase diagram:} We begin by analyzing the phase diagram of the model \eqref{adiabatic} in the adiabatic limit. 
Now determining the ground state reduces to minimizing the energy over both phonon and flux configurations. 
This problem is greatly simplified by invoking two  Lieb's theorems which are based on reflection positivity \cite{PhysRevLett.73.2158,PhysRevLett.107.066801,SupMat}. The relevant symmetry required by Lieb's theorems is the reflection symmetry with mirror planes $\mathcal{M}$ bisecting the $z$-type of bonds (namely, $y\to -y$ under the reflection). Owing to these theorems, the ground state lies in the zero flux sector and  the most general configuration of the phonon fields $X_{\langle ij\rangle}$ in the ground state  is symmetric under $\mathcal{M}$. Importantly, Lieb’s theorems guarantee zero flux of the total hopping amplitude $t_{\langle ij\rangle}=\hat{u}_{\langle ij\rangle}(a_{\mu}+\hat{X}_{\langle ij\rangle})$, not necessarily zero flux of the gauge field $\hat{u}_{\langle ij\rangle}$ itself. Consequently, two scenarios arise: (1) If all the phonon fields $\hat{X}_{\langle ij\rangle}$ are small enough such that all hoppings $a_{\mu}+\hat{X}_{\langle ij\rangle}$ are positive, then the  $\hat{u}_{\langle ij\rangle}$ flux is also zero; (2) If some of the $\hat{X}_{\langle ij\rangle}$ are negative enough to induce $\pi$ flux in the hoppings $a_{\mu}+\hat{X}_{\langle ij\rangle}$ on certain plaquettes, then the corresponding  $\hat{u}_{\langle ij\rangle}$  flux should also be $\pi$  to preserve zero net flux of $t_{\langle ij\rangle}$.

Here we go beyond Lieb's theorems and prove that the flux of $u_{\langle ij\rangle}$ is actually zero in the ground state sector. This result is established through a proof-by-contradiction. If the flux of $u_{\langle ij\rangle}$ is $\pi$ around a plaquette $p$ in the ground state sector: $\Pi_{\langle ij\rangle\in p}u_{\langle ij\rangle}=-1$, then some of the phonon fields $X_{\langle ij\rangle}$ and $a_{\mu}+X_{\langle ij\rangle}$ $(\langle ij\rangle\in p)$ must be negative, since the total flux of $\Pi_{\langle ij\rangle\in p}\left(u_{\langle ij\rangle}(a_{\mu}+X_{\langle ij\rangle})\right)$ is required to be zero by Lieb's theorems. Then we can find new phonon and gauge field configurations on these bonds with strictly lower energy: $\Tilde{X}_{\langle ij\rangle}=-2a_{\mu}-X_{\langle ij\rangle}$, $\quad \Tilde{u}_{\langle ij\rangle}=-u_{\langle ij\rangle}$.  This configuration preserves the hopping amplitudes: $\tilde{t}_{\langle ij\rangle}=(a_{\mu}+\tilde{X}_{\langle ij\rangle})\tilde{u}_{\langle ij\rangle}=(a_{\mu}+ X_{\langle ij\rangle})u_{\langle ij\rangle}=t_{\langle ij\rangle}$, and thus the fermion energy is unchanged.  However, the new phonon configuration yields a strictly lower phonon potential energy,  since 
$|\tilde{X}_{\braket{ij}}|^2-|X_{\braket{ij}}|^2=4a_{\mu}(a_{\mu}+X_{\braket{ij}})<0$. 
This process can be iterated until all plaquettes satisfy $\Pi_{\langle ij\rangle\in p}\tilde{u}_{\braket{ij}}=1$, which has strictly lower energy than any state with $\pi$-flux of $u_{\braket{ij}}$, which contradicts our initial assumption. This completes the proof. Hence, we set all the $\mathbb{Z}_2$ gauge fields $\hat{u}_{\langle ij \rangle}=1$ in the following discussions. A direct implication of our proof is the effective $\tilde{J}_{\mu} = a_{\mu} + X_{\mu}\geq0$ in the ground state sector.

Furthermore, Lieb's theorems do not address whether the ground state exhibits spontaneous symmetry breaking. Consequently, there are two possible scenarios in principle:

(1) For sufficiently small $\lambda$, it is expected that the phonon fields $X_{\langle ij\rangle}$ preserve translation symmetries; that is, $X_{\langle ij\rangle \in \mu}$ satisfies $X_{\langle ij\rangle \in \mu} = X_{\mu}$, resulting in an itinerant fermion spectrum that resembles the pure Kitaev honeycomb model with anisotropic couplings $\tilde{J}_{\mu} = a_{\mu} + X_{\mu}$. 
However, in this scenario, the ground state always remains a gapless spin liquid with a single Dirac cone and can never enter a gapped phase.  
This is because the hopping amplitudes satisfy $\tilde J_x=\tilde J_y$ in the ground state sector, as follows directly from Lieb's theorems \cite{PhysRevLett.73.2158,PhysRevLett.107.066801,SupMat} together with the zero flux condition for $\hat{u}_{\langle ij\rangle}$ we proved above, and are bounded within $0 \leq \tilde{J}_z/\tilde{J}_x < 1$ for any $\lambda$, preventing the system from reaching the anisotropic limit $|\tilde{J}_z| > \tilde{J}_x+\tilde{J}_y$ where the Kitaev model becomes gapped. $\tilde{J}_z$ can never be larger than  $\tilde{J}_x$, since the Dirac cone can only move along the $k_x$-axis with $k_x \in [\pi, \frac{4\pi}{3})$, where $\frac{\tilde{J}_z}{\tilde{J}_x}\rightarrow 1$ as $k_x$ approaches $\frac{4\pi}{3}$. However, the endpoint $k_x=\frac{4\pi}{3}$, which would imply isotropic $\tilde{J}_x=\tilde{J}_y=\tilde{J}_z$, is not attainable, as the energy minimum results in a $C_3$ symmetric $X_{x}=X_{y}=X_{z}$ phonon configuration, which contradicts with the isotropic $\tilde{J}_{\mu}$ coupling.



(2) For a sufficiently large $\lambda$ beyond a critical value $\lambda_c$, the phonon fields $X_{\langle ij\rangle}$ break the translation symmetry. Driven by the Peierls instability, the momentum of this $X_{\langle ij\rangle}$ configuration is expected to match the intra-valley scattering momentum of the Dirac cone, when the coupling $\lambda$ just exceeds the critical $\lambda_c$. So, the ground state typically has an incommensurate VBS order 
coexisting with $\mathbb{Z}_2$ topological order (in the sense the collective Goldstone mode is neglected) in this phase.
 
As a result, the phase diagram is governed by spontaneous breaking of the translation symmetry as the spin-phonon coupling $\lambda$ increases. We obtain the global quantum phase diagram through large-scale numerical simulations, as is illustrated in Fig.~\ref{AdiaPhaseDiag}(a), which agrees well with the previous theoretical analyses. In particular, all the phases depicted in Fig.~\ref{AdiaPhaseDiag} are quantum spin liquids. This is due to the persistence of exact anomalous 1-form symmetry and deconfined fermions in all phases \cite{PhysRevB.67.245316,PhysRevB.101.115113}. 
Although the strong-coupling columnar $\text{VBS}$ topological order phase in Fig.~\ref{AdiaPhaseDiag}(a) is compelling, our primary focus is on the QCP between the gapless spin liquid and the incommensurate $\text{VBS}$ spin liquid phases. 
Interestingly, a vortex in the incommensurate $\text{VBS}$ spin liquid can trap a Majorana zero mode~\cite{fu2008superconducting}, which in this context behaves as an Ising anyon. 
As shown in Fig.~\ref{AdiaPhaseDiag}(b), this phase transition is continuous in the adiabatic limit.
Through field theory analysis, we demonstrate that there is an emergent $\mathcal{N}=2$ spacetime SUSY at these QCPs, when the quantum fluctuations of the phonon fields are further included. 
Additionally, we establish that the transitions into commensurate $\text{VBS}$ topological order phases belong to the same supersymmetric universality class, provided that their commensurability 
exceed four. 

{\bf 
Field theory in QCPs:} We now develop the field theory for the quantum critical points between the gapless spin liquid and the topological ordered $\text{VBS}$. 
Given that the gauge fields $\hat{u}_{\langle ij\rangle}$ are always conserved, only the itinerant fermions $\hat{c}_i$ and the phonon fields are dynamic degrees of freedom, so we fix the gauge $\hat{u}_{\langle ij\rangle}=1$ hereafter \footnote{Strictly speaking, the physical state is an equal weight superposition of all the states connected to the gauge-fixed $\hat{u}_{ij}=1$ wavefunction $|\psi\rangle$  by $\mathbb{Z}_2$ gauge transformations \cite{PhysRevLett.102.217202}. Nevertheless, the behavior of physical/gauge-invariant quantities---such as emergent SUSY or correlation functions---are the same as those obtained in the field theory developed below by fixing the gauge}. Here, we incorporate quantum fluctuations of the phonon field by taking a finite yet sufficiently large phonon mass $m$, rendering the phonon dynamics a perturbative factor that does not qualitatively alter the phase diagram and the ground state still lies in the zero $\mathbb{Z}_2$ flux sector. This assumption is supported by our 
perturbative calculations at finite phonon frequency \cite{SupMat}, and the finite $\omega_{ph}$ phase diagram is consistent with Fig. \ref{AdiaPhaseDiag}(c).


The critical field theory contains three parts: $S=S_f+S_b+S_\text{int}$, where we should retain all the relevant terms allowed for symmetry. $S_f=\int d \tau d^{2} x{\psi^\dagger}( \partial_{\tau}-iv_x\partial_x\sigma_y-iv_y\partial_y\sigma_x )\psi$ is the action of a Dirac cone obtained by expanding the Majorana fermions $\hat{c}_i$ at the band touching point $\mathbf{\tilde{K}}$ \footnote{Here $\mathbf{\tilde{K}}$ does not mean the corner of the Brillouin zone, since we have included anisotropic hoppings of the Majorana fermions.}: $(\hat{c}_{A}(\mathbf{r}_i),\hat{c}_{B}(\mathbf{r}_i))^T\approx\hat{\psi}(x)e^{i\mathbf{\tilde{K}}\cdot \mathbf{r}_i}+\text{h.c.}$. Here we take the two nearest-neighbor sites in a $z$-bond as a unit cell labeled by $\mathbf{r}_i$.

$S_b$ describes the quantum fluctuation of the order parameter $\phi(\vec{x},\tau)$: $S_b=\int d \tau d^{2} x\left|\partial_{\tau} \phi\right|^{2}+\sum_{i=x,y}v_{b,i}^{2}|\partial_i \phi|^{2}+V(\phi,\phi^*)$, where $\phi(\vec{x},\tau)$ is a complex bosonic field with momentum $-2\mathbf{\tilde K}$ under the lattice translation: $\hat{T}_{\mathbf{a}_i}\phi(\vec{x},\tau)\hat{T}^{-1}_{\mathbf{a}_i}=e^{-2i\mathbf{\tilde{K}}\cdot\mathbf{a}_i}\phi(\vec{x},\tau)$, where $\mathbf{a}_{i=1,2}$ are the unit vectors of the honeycomb lattice.
If the ordered VBS phase is incommensurate, then the translation symmetry becomes an emergent $U(1)$ symmetry and the symmetry allowed potential $V(\phi,\phi^*)$ can only depend on the module of $\phi(\vec{x},\tau)$: $V(\phi,\phi^*)=u\int d\tau d^2x |\phi(\vec{x},\tau)|^{4}$, where we tune the mass term of $\phi(\vec{x},\tau)$ to be zero, since we are considering a critical point. 
On the other hand, if the VBS order is commensurate with the lattice, which means that $2n\mathbf{\tilde{K}}\equiv0$ mod reciprocal momentum, then the lattice symmetry also allows for an additional term $r_n\int d\tau d^2x[\phi^n+(\phi^*)^n]$ in the potential $V(\phi,\phi^*)$. In addition, although $i\phi^*\partial_{\tau}\phi$ is typically allowed in the complex boson kinetic part, it is prohibited by an inversion symmetry $\mathcal{I}$ here: $\psi^T(\vec{x},\tau)\rightarrow\bar\psi(-\vec{x},\tau)\sigma_y,\phi(\vec{x},\tau)\rightarrow\phi^{*}(-\vec{x},\tau)$. Here $\phi$ becomes its complex conjugate under inversion since it carries finite momentum. 
Finally, the symmetry-allowed interaction $S_\text{int}$ is a Yukawa-type coupling: $S_{\text{int}}=g\int d\tau d^2x\left(\phi \psi^{T} \sigma_y \psi+\text{h.c.}\right)$.

Microscopically, we can derive the Yukawa coupling from the SSH coupling by relating the bosonic field $\phi(\vec{x},\tau)$ to the phonon field $\hat{X}_{\langle ij\rangle\in\mu}$ on the lattice. In each unit cell, there are three phonon fields labeled by $\mu$ (corresponding to the $\mu$-type bonds in the Kitaev interaction). Crucially, we only retain those phonon modes with momenta near $\pm2\mathbf{\tilde{K}}$ at the critical point, since only these modes couple with the low-energy fermions. For each bond type $\mu$, the field $\hat{X}_{\langle ij\rangle\in \mu}$ can therefore be approximated as: $\hat{X}_{\langle ij\rangle\in \mu}\approx e^{-2i\mathbf{\tilde{K}}\cdot \mathbf{r}_i}\hat{\phi}^{\mu}(\vec{x})+\text{h.c.}$, where $\hat{\phi}^{\mu}(\vec{x})$ is a slowly varying field compared to the lattice constant (where $\mathbf{r}_i$ denotes the position of site $i$ in the A sublattice) and $\hat{\phi}(\vec{x})$ is the linear combination of $\hat{\phi}^{\mu}$: $\sum_{\mu}e^{i\mathbf{\tilde{K}}\cdot \mathbf{e}_\mu}\hat{\phi}^{\mu}(\vec{x})=\hat{\phi}(\vec{x})$, where  $\mathbf{e}_x=(-\frac{1}{2},-\frac{\sqrt{3}}{2})$, $\mathbf{e}_y=(\frac{1}{2},-\frac{\sqrt{3}}{2})$ and $\mathbf{e}_z=(0,0)$ since it is the intra-unit cell vector. 
Taking this continuum limit, the SSH spin-phonon coupling reduces to the Yukawa coupling described by the action $S_{\text{int}}$. 
Furthermore, the inversion symmetry $\mathcal{I}$ of the continuum theory directly inherits from the lattice inversion symmetry of the original SSH-Kitaev model.

{\bf Emergent SUSY at fractionalized QCPs:} In this section, we investigate the emergent supersymmetry at the fractionalized QCP described by the above critical field theory $S$. 
We begin with the phase transition between the gapless spin liquid and the incommensurate $\text{VBS}$ topological order, where the potential $V(\phi,\phi^*)$ depends solely on the modulus of $\phi(\vec{x},\tau)$. Renormalization group calculations \cite{doi:10.1126/science.1248253,PhysRevB.76.075103,PhysRevLett.114.237001,SupMat} indicate that the critical field theory $S$ flows towards a supersymmetric fixed point, specifically the $\mathcal{N}=2$  supersymmetric Wess-Zumino model. 
This emergent SUSY has also been corroborated through sign-problem-free quantum Monte Carlo simulations, achieved by discretizing the single Dirac cone on a lattice using `SLAC' fermions with long-range hoppings \cite{doi:10.1126/sciadv.aau1463}.

Then we move to the fractionalized QCP between the gapless spin liquid and the commensurate $\text{VBS}$ topological order. 
The additional term, $r_n\int d\tau d^2x(\phi^n+(\phi^*)^n)$, can be treated as a perturbation to the supersymmetric Wess-Zumino fixed point. 
Owing to the supersymmetry, the scaling dimension of the field $\phi^n$ is exactly known as $\frac{2n}{3}$ \cite{seiberg1994powerholomorphyexact}. As a result, if the period $n$ of the VBS order ($2n\mathbf{\tilde{K}}\equiv0$ mod reciprocal momentum) satisfies $n\ge 5$, then $\phi^n$ is irrelevant in the renormalization group sense and the fixed point remains the supersymmetric Wess-Zumino model. A schematic phase diagram with a finite phonon frequency is illustrated in Fig. \ref{AdiaPhaseDiag}(c).

More precisely, our fractionalized QCP should belong to the $\text{SUSY}^*$ universality class since the constituent fermions come from the fractionalization of physical spin operators. 
This $\text{SUSY}^*$ not only shares the general properties of fractionalized QCPs, such as the appearance of multiple copies of the standard SUSY spectrum when placed on a torus \cite{PhysRevLett.117.210401,PhysRevLett.125.257202}, but also exhibits intrinsically new  physical consequences arising from SUSY which are absent in conventional unfractionalized SUSY QCPs. 
In particular, while SUSY usually maps local bosonic to local fermionic operators, here it relates the local bosonic field $\phi(\vec{x},\tau)$ to a nonlocal fermionic operator dressed with a $\mathbb{Z}_2$ gauge string connected to infinity. 
This arises because physical operators and states must be $\mathbb{Z}_2$ gauge-invariant, whereas the itinerate Majorana fermions $\hat{c}_i$ alone are not. 
This ``local-nonlocal'' SUSY correspondence is also manifest in correlation functions. 
Emergent SUSY enforces identical anomalous dimensions $\eta_f=\eta_b=\frac{1}{3}$. 
Although $\eta_b$ appears in a local phonon correlation, $\eta_f$ is only revealed in a gauge-invariant nonlocal string operator $\langle \Pi_{\langle lm\rangle\in L}(\tau^{\mu}_l\tau^{\mu}_m)\rangle\propto \frac{1}{|r_i-r_{i^{\prime}}|^{2+\eta_f}}$, where $L$ is a path connecting sites $i$ and $i^{\prime}$ and $\mu$ is the label of the bond $\langle lm\rangle$ in the Kitaev interaction. This string correlator is precisely the gauge-invariant version of the dynamical Majorana two-point function dressed by a $\mathbb{Z}_2$ gauge string: $\langle \hat{c}_i(\hat{u}_{\langle ij\rangle}\hat{u}_{\langle jk\rangle}...\hat{u}_{\langle h^{\prime}i^{\prime}\rangle}) \hat{c}_{i^{\prime}}\rangle$. This is an intrinsically fractionalized feature with no analogue in unfractionalized QCPs, where $\eta_f$ is directly visible in a local fermion correlator.

Finally, the appearance of $\eta_f$ in a nonlocal string operator is not a fine-tuned property of our Kitaev-SSH model, but  rather a general property of the $\text{SUSY}^*$ fractionalized QCP between two $\mathbb{Z}_2$ spin liquid phases with deconfined $\mathbb{Z}_2$ gauge fields, although the microscopic form of the string operator may become more extended or ``thicker'' in a perturbed model (e.g. with the inclusion of a small Heisenberg term). 



{\bf Experimental signatures of SUSY fixed points:} Possible experimental signatures of the supersymmetric QCP between the gapless spin liquid and topological order phases are provided by universal scaling exponents of physical quantities.  
Since the system is an electric insulator, a typical physical quantity of transport is the longitudinal thermal conductivity, which can be obtained from the Kubo formula through analytical continuation: $\kappa^{ii}(\omega)=\kappa_{\text{Kubo}}^{ii}(\omega_n)|_{\omega_n\rightarrow-i\omega+\delta}$, where $\kappa_{\text{Kubo}}^{ii}(i\omega_n)=\frac{1}{\omega_n} \langle J_{Q}^{i}(\omega_n) J^i_{Q}(-\omega_n))\rangle$ with $\omega_n=2\pi nT$ being the Matsubara frequency and $J_{Q}^{i}(\omega_n)$ is the heat current operator in the spatial direction $i$. 
In the high-frequency regime $\hbar\omega\gg k_BT$ and neglecting the contributions of gapped flux excitations, we can show that $\kappa^{ii}(\omega)$ scales as: $\kappa^{ii}(\omega)\propto (i\omega)^{2-\Delta} T^{\Delta}$ \cite{SupMat} using the operator product expansion (OPE) method in \cite{PhysRevB.90.245109}, where $\Delta=3-\frac{1}{\nu}\approx 1.9098$ is the scaling dimension of the bosonic field $|\phi|^2(\vec{x},\tau)$ at the critical point \cite{PhysRevLett.115.051601,PhysRevLett.116.100402}. In addition, the zero-temperature dynamical shear viscosity $\eta(\omega,T=0)$ also takes a universal form $\eta(\omega,T=0)=\eta_{\infty}\omega^2\hbar$ at the SUSY critical point \cite{PhysRevLett.116.100402}, where $\eta_{\infty}\approx 5.68\times10^{-3}$. 

{\bf Discussions and concluding remarks:} In conclusion, we have shown that coupling a Kitaev quantum spin liquid to phonons via an SSH-type interaction yields a rich sequence of fractionalized phases and continuous transitions from a Dirac QSL to incommensurate/commensurate VBS coexisting with $\mathbb{Z}_2$ topological order. 
Using Lieb's theorems, large-scale numerical computations, and low-energy field theory, we have demonstrated that quantum phonon fluctuations drive these fractionalized quantum critical points to an emergent $\mathcal{N}=2$ spacetime $\text{SUSY}^*$ fixed point. 
This provides a rare and concrete lattice realization of SUSY in a strongly correlated setting with deconfined fractionalized particles. We further identify universal signatures in thermal transport and shear viscosity that can serve as experimental probes in Kitaev-like materials \cite{PhysRevLett.102.017205,annurev:/content/journals/10.1146/annurev-conmatphys-031115-011319,annurev:/content/journals/10.1146/annurev-conmatphys-033117-053934,wen2019experimental,TREBST20221} with strong spin-lattice coupling, highlighting spin-phonon interactions as a promising route for engineering emergent supersymmetry in two dimensions.

{\bf Acknowledgments:} We sincerely thank Siddharth Parameswaran for helpful discussions. This work is supported in part by the NSFC under Grant Nos. 12347107 and 12334003 (Z.W., Z.-Q.W., and H.Y.), MOSTC under Grant No. 2021YFA1400100 (H.Y.), and the New Cornerstone Science Foundation through the Xplorer Prize (H.Y.). Z.W. acknowledges the support in part from the Shuimu fellowship at Tsinghua University and the EPSRC under grant EP/X030881/1. The Flatiron Institute is a division
of the Simons Foundation.  The work of S.-K.J. is supported by a start-up fund at Tulane University.

\bibliography{bib}

@article{PhysRevB.83.180412,
  title = {Exact chiral spin liquid with stable spin Fermi surface on the kagome lattice},
  author = {Chua, Victor and Yao, Hong and Fiete, Gregory A.},
  journal = {Phys. Rev. B},
  volume = {83},
  issue = {18},
  pages = {180412},
  numpages = {4},
  year = {2011},
  month = {May},
  publisher = {American Physical Society},
  doi = {10.1103/PhysRevB.83.180412},
  url = {https://link.aps.org/doi/10.1103/PhysRevB.83.180412}
}

@article{Wu2009GammaMatrix,
  title   = {A $\Gamma$-matrix generalization of the Kitaev model},
  author  = {Wu, Congjun and Arovas, Daniel and Hung, Hsiang-Hsuan},
  journal = {Phys. Rev. B},
  volume  = {79},
  pages   = {134427},
  year    = {2009},
  doi     = {10.1103/PhysRevB.79.134427}
}

@article{PhysRevLett.99.196805,
  title = {Edge Solitons of Topological Insulators and Fractionalized Quasiparticles in Two Dimensions},
  author = {Lee, Dung-Hai and Zhang, Guang-Ming and Xiang, Tao},
  journal = {Phys. Rev. Lett.},
  volume = {99},
  issue = {19},
  pages = {196805},
  numpages = {4},
  year = {2007},
  month = {Nov},
  publisher = {American Physical Society},
  doi = {10.1103/PhysRevLett.99.196805},
  url = {https://link.aps.org/doi/10.1103/PhysRevLett.99.196805}
}

@article{Chulliparambil2020SixteenfoldMicroscopic,
  title   = {Microscopic models for Kitaev's sixteenfold way of anyon theories},
  author  = {Chulliparambil, Sreejith and Seifert, Urban F. P. and Vojta, Matthias and Janssen, Lukas and Tu, Hong-Hao},
  journal = {Phys. Rev. B},
  volume  = {102},
  pages   = {201111},
  year    = {2020},
  doi     = {10.1103/PhysRevB.102.201111}
}

@article{Jin2023SixteenfoldExactlySolvable,
  title   = {Exactly solvable models for Kitaev's sixteen-fold way},
  author  = {Jin, Jin-Tao and Miao, Jian-Jian and Zhou, Yi},
  journal = {SciPost Phys.},
  volume  = {14},
  number  = {5},
  pages   = {087},
  year    = {2023},
  doi     = {10.21468/SciPostPhys.14.5.087}
}

@Article{Eck2025BraidedFusionGeneralizations,
	title={{Generalizations of Kitaev’s honeycomb model from braided fusion categories}},
	author={Luisa Eck and Paul Fendley},
	journal={SciPost Phys.},
	volume={18},
	pages={170},
	year={2025},
	publisher={SciPost},
	doi={10.21468/SciPostPhys.18.6.170},
	url={https://scipost.org/10.21468/SciPostPhys.18.6.170},
}

@article{Baskaran2008SpinS,
  title   = {Spin-$S$ Kitaev model: Classical ground states, order from disorder, and exact correlation functions},
  author  = {Baskaran, G. and Sen, Diptiman and Shankar, R.},
  journal = {Phys. Rev. B},
  volume  = {78},
  pages   = {115116},
  year    = {2008},
  doi     = {10.1103/PhysRevB.78.115116}
}

@article{wen2019experimental,
  title={Experimental identification of quantum spin liquids},
  author={Wen, Jinsheng and Yu, Shun-Li and Li, Shiyan and Yu, Weiqiang and Li, Jian-Xin},
  journal={npj Quantum Materials},
  volume={4},
  number={1},
  pages={12},
  year={2019},
  publisher={Nature Publishing Group UK London},
doi={10.1038/s41535-019-0151-6},
url={https://journals.aps.org/prl/pdf/10.1103/c8n5-l11j}
}

@article{fu2008superconducting,
  title = {Superconducting Proximity Effect and Majorana Fermions at the Surface of a Topological Insulator},
  author = {Fu, Liang and Kane, C. L.},
  journal = {Phys. Rev. Lett.},
  volume = {100},
  issue = {9},
  pages = {096407},
  numpages = {4},
  year = {2008},
  month = {Mar},
  publisher = {American Physical Society},
  doi = {10.1103/PhysRevLett.100.096407},
  url = {https://link.aps.org/doi/10.1103/PhysRevLett.100.096407}
}

@article{PhysRevLett.114.157202,
  title = {Weyl Spin Liquids},
  author = {Hermanns, M. and O'Brien, K. and Trebst, S.},
  journal = {Phys. Rev. Lett.},
  volume = {114},
  issue = {15},
  pages = {157202},
  numpages = {5},
  year = {2015},
  month = {Apr},
  publisher = {American Physical Society},
  doi = {10.1103/PhysRevLett.114.157202},
  url = {https://link.aps.org/doi/10.1103/PhysRevLett.114.157202}
}

@article{PhysRevB.79.024426,
  title = {Exactly solvable Kitaev model in three dimensions},
  author = {Mandal, Saptarshi and Surendran, Naveen},
  journal = {Phys. Rev. B},
  volume = {79},
  issue = {2},
  pages = {024426},
  numpages = {8},
  year = {2009},
  month = {Jan},
  publisher = {American Physical Society},
  doi = {10.1103/PhysRevB.79.024426},
  url = {https://link.aps.org/doi/10.1103/PhysRevB.79.024426}
}

@article{PhysRevLett.107.087205,
  title = {Fermionic Magnons, Non-Abelian Spinons, and the Spin Quantum Hall Effect from an Exactly Solvable Spin-$1/2$ Kitaev Model with SU(2) Symmetry},
  author = {Yao, Hong and Lee, Dung-Hai},
  journal = {Phys. Rev. Lett.},
  volume = {107},
  issue = {8},
  pages = {087205},
  numpages = {5},
  year = {2011},
  month = {Aug},
  publisher = {American Physical Society},
  doi = {10.1103/PhysRevLett.107.087205},
  url = {https://link.aps.org/doi/10.1103/PhysRevLett.107.087205}
}

@article{hongchiral,
  title = {Exact Chiral Spin Liquid with Non-Abelian Anyons},
  author = {Yao, Hong and Kivelson, Steven A.},
  journal = {Phys. Rev. Lett.},
  volume = {99},
  issue = {24},
  pages = {247203},
  numpages = {4},
  year = {2007},
  month = {Dec},
  publisher = {American Physical Society},
  doi = {10.1103/PhysRevLett.99.247203},
  url = {https://link.aps.org/doi/10.1103/PhysRevLett.99.247203}
}

@article{PhysRevB.76.075103,
  title = {Emergence of supersymmetry at a critical point of a lattice model},
  author = {Lee, Sung-Sik},
  journal = {Phys. Rev. B},
  volume = {76},
  issue = {7},
  pages = {075103},
  numpages = {6},
  year = {2007},
  month = {Aug},
  publisher = {American Physical Society},
  doi = {10.1103/PhysRevB.76.075103},
  url = {https://link.aps.org/doi/10.1103/PhysRevB.76.075103}
}

@article{PhysRevB.94.085134,
  title = {Transition from the ${\mathbb{Z}}_{2}$ spin liquid to antiferromagnetic order: Spectrum on the torus},
  author = {Whitsitt, Seth and Sachdev, Subir},
  journal = {Phys. Rev. B},
  volume = {94},
  issue = {8},
  pages = {085134},
  numpages = {16},
  year = {2016},
  month = {Aug},
  publisher = {American Physical Society},
  doi = {10.1103/PhysRevB.94.085134},
  url = {https://link.aps.org/doi/10.1103/PhysRevB.94.085134}
}

@article{PhysRevLett.102.017205,
  title = {Mott Insulators in the Strong Spin-Orbit Coupling Limit: From Heisenberg to a Quantum Compass and Kitaev Models},
  author = {Jackeli, G. and Khaliullin, G.},
  journal = {Phys. Rev. Lett.},
  volume = {102},
  issue = {1},
  pages = {017205},
  numpages = {4},
  year = {2009},
  month = {Jan},
  publisher = {American Physical Society},
  doi = {10.1103/PhysRevLett.102.017205},
  url = {https://link.aps.org/doi/10.1103/PhysRevLett.102.017205}
}

@article{TREBST20221,
title = {Kitaev materials},
journal = {Physics Reports},
volume = {950},
pages = {1-37},
year = {2022},
issn = {0370-1573},
doi = {https://doi.org/10.1016/j.physrep.2021.11.003},
url = {https://www.sciencedirect.com/science/article/pii/S0370157321004051},
author = {Simon Trebst and Ciarán Hickey}
}

@article{annurev:/content/journals/10.1146/annurev-conmatphys-033117-053934,
   author = "Hermanns, M. and Kimchi, I. and Knolle, J.",
   title = "Physics of the Kitaev Model: Fractionalization, Dynamic Correlations, and Material Connections", 
   journal= "Annual Review of Condensed Matter Physics",
   year = "2018",
   volume = "9",
   number = "Volume 9, 2018",
   pages = "17-33",
   doi = "https://doi.org/10.1146/annurev-conmatphys-033117-053934",
   url = "https://www.annualreviews.org/content/journals/10.1146/annurev-conmatphys-033117-053934",
   publisher = "Annual Reviews",
   issn = "1947-5462",
   type = "Journal Article",
  }

@article{PhysRevLett.133.236504,
  title = {Exact Deconfined Gauge Structures in the Higher-Spin {Yao-Lee} Model: A Quantum Spin-Orbital Liquid with Spin Fractionalization and Non-Abelian Anyons},
  author = {Wu, Zhengzhi and Zhang, Jing-Yun and Yao, Hong},
  journal = {Phys. Rev. Lett.},
  volume = {133},
  issue = {23},
  pages = {236504},
  numpages = {7},
  year = {2024},
  month = {Dec},
  publisher = {American Physical Society},
  doi = {10.1103/PhysRevLett.133.236504},
  url = {https://link.aps.org/doi/10.1103/PhysRevLett.133.236504}
}

@article{PhysRevResearch.6.043273,
  title = {Supersymmetry on the lattice: Geometry, topology, and flat bands},
  author = {Roychowdhury, Krishanu and Attig, Jan and Trebst, Simon and Lawler, Michael J.},
  journal = {Phys. Rev. Res.},
  volume = {6},
  issue = {4},
  pages = {043273},
  numpages = {38},
  year = {2024},
  month = {Dec},
  publisher = {American Physical Society},
  doi = {10.1103/PhysRevResearch.6.043273},
  url = {https://link.aps.org/doi/10.1103/PhysRevResearch.6.043273}
}

@article{PhysRevLett.130.156701,
  title = {${Z}_{2}$ Spin Liquids in the Higher Spin-${S}$ Kitaev Honeycomb Model: An Exact Deconfined ${Z}_{2}$ Gauge Structure in a Nonintegrable Model},
  author = {Ma, Han},
  journal = {Phys. Rev. Lett.},
  volume = {130},
  issue = {15},
  pages = {156701},
  numpages = {6},
  year = {2023},
  month = {Apr},
  publisher = {American Physical Society},
  doi = {10.1103/PhysRevLett.130.156701},
  url = {https://link.aps.org/doi/10.1103/PhysRevLett.130.156701}
}

@article{PhysRevD.68.065003,
  title = {Quantum order from string-net condensations and the origin of light and massless fermions},
  author = {Wen, Xiao-Gang},
  journal = {Phys. Rev. D},
  volume = {68},
  issue = {6},
  pages = {065003},
  numpages = {25},
  year = {2003},
  month = {Sep},
  publisher = {American Physical Society},
  doi = {10.1103/PhysRevD.68.065003},
  url = {https://link.aps.org/doi/10.1103/PhysRevD.68.065003}
}

@article{annurev:/content/journals/10.1146/annurev-conmatphys-031115-011319,
   author = "Rau, Jeffrey G. and Lee, Eric Kin-Ho and Kee, Hae-Young",
   title = "Spin-Orbit Physics Giving Rise to Novel Phases in Correlated Systems: Iridates and Related Materials", 
   journal= "Annual Review of Condensed Matter Physics",
   year = "2016",
   volume = "7",
   number = "Volume 7, 2016",
   pages = "195-221",
   doi = "https://doi.org/10.1146/annurev-conmatphys-031115-011319",
   url = "https://www.annualreviews.org/content/journals/10.1146/annurev-conmatphys-031115-011319",
   publisher = "Annual Reviews",
   issn = "1947-5462",
   type = "Journal Article",
  }

@article{
doi:10.1073/pnas.1806338115,
author = {Snir Gazit  and Fakher F. Assaad  and Subir Sachdev  and Ashvin Vishwanath  and Chong Wang },
title = {Confinement transition of Z2 gauge theories coupled to massless fermions: Emergent quantum chromodynamics and SO(5) symmetry},
journal = {Proceedings of the National Academy of Sciences},
volume = {115},
number = {30},
pages = {E6987-E6995},
year = {2018},
doi = {10.1073/pnas.1806338115},
URL = {https://www.pnas.org/doi/abs/10.1073/pnas.1806338115}}

@article{Vojta_2018,
doi = {10.1088/1361-6633/aab6be},
url = {https://dx.doi.org/10.1088/1361-6633/aab6be},
year = {2018},
month = {may},
publisher = {IOP Publishing},
volume = {81},
number = {6},
pages = {064501},
author = {Vojta, Matthias},
title = {Frustration and quantum criticality},
journal = {Reports on Progress in Physics}
}

@article{PhysRevLett.117.210401,
  title = {Universal Signatures of Quantum Critical Points from Finite-Size Torus Spectra: A Window into the Operator Content of Higher-Dimensional Conformal Field Theories},
  author = {Schuler, Michael and Whitsitt, Seth and Henry, Louis-Paul and Sachdev, Subir and L\"auchli, Andreas M.},
  journal = {Phys. Rev. Lett.},
  volume = {117},
  issue = {21},
  pages = {210401},
  numpages = {6},
  year = {2016},
  month = {Nov},
  publisher = {American Physical Society},
  doi = {10.1103/PhysRevLett.117.210401},
  url = {https://link.aps.org/doi/10.1103/PhysRevLett.117.210401}
}

@article{
doi:10.1126/science.1212207,
author = {Sergei V. Isakov  and Roger G. Melko  and Matthew B. Hastings },
title = {Universal Signatures of Fractionalized Quantum Critical Points},
journal = {Science},
volume = {335},
number = {6065},
pages = {193-195},
year = {2012},
doi = {10.1126/science.1212207},
URL = {https://www.science.org/doi/abs/10.1126/science.1212207}}

@article{PhysRevLett.125.257202,
  title = {Fractionalized Fermionic Quantum Criticality in Spin-Orbital Mott Insulators},
  author = {Seifert, Urban F. P. and Dong, Xiao-Yu and Chulliparambil, Sreejith and Vojta, Matthias and Tu, Hong-Hao and Janssen, Lukas},
  journal = {Phys. Rev. Lett.},
  volume = {125},
  issue = {25},
  pages = {257202},
  numpages = {7},
  year = {2020},
  month = {Dec},
  publisher = {American Physical Society},
  doi = {10.1103/PhysRevLett.125.257202},
  url = {https://link.aps.org/doi/10.1103/PhysRevLett.125.257202}
}

@article{PhysRevLett.114.090404,
  title = {Emergent Supersymmetry at the Ising--Berezinskii-Kosterlitz-Thouless Multicritical Point},
  author = {Huijse, Liza and Bauer, Bela and Berg, Erez},
  journal = {Phys. Rev. Lett.},
  volume = {114},
  issue = {9},
  pages = {090404},
  numpages = {5},
  year = {2015},
  month = {Mar},
  publisher = {American Physical Society},
  doi = {10.1103/PhysRevLett.114.090404},
  url = {https://link.aps.org/doi/10.1103/PhysRevLett.114.090404}
}

@article{PhysRevLett.115.166401,
  title = {Emergent Supersymmetry from Strongly Interacting Majorana Zero Modes},
  author = {Rahmani, Armin and Zhu, Xiaoyu and Franz, Marcel and Affleck, Ian},
  journal = {Phys. Rev. Lett.},
  volume = {115},
  issue = {16},
  pages = {166401},
  numpages = {5},
  year = {2015},
  month = {Oct},
  publisher = {American Physical Society},
  doi = {10.1103/PhysRevLett.115.166401},
  url = {https://link.aps.org/doi/10.1103/PhysRevLett.115.166401}
}

@article{Ponte_2014,
doi = {10.1088/1367-2630/16/1/013044},
url = {https://dx.doi.org/10.1088/1367-2630/16/1/013044},
year = {2014},
month = {jan},
publisher = {IOP Publishing},
volume = {16},
number = {1},
pages = {013044},
author = {Ponte, Pedro and Lee, Sung-Sik},
title = {Emergence of supersymmetry on the surface of three-dimensional topological insulators},
journal = {New Journal of Physics}
}

@article{KITAEV20062,
title = {Anyons in an exactly solved model and beyond},
journal = {Annals of Physics},
volume = {321},
number = {1},
pages = {2-111},
year = {2006},
note = {January Special Issue},
issn = {0003-4916},
doi = {https://doi.org/10.1016/j.aop.2005.10.005},
url = {https://www.sciencedirect.com/science/article/pii/S0003491605002381},
author = {Alexei Kitaev}
}

@article{PhysRevB.10.4637,
  title = {Peierls instability in Heisenberg chains},
  author = {Pytte, E.},
  journal = {Phys. Rev. B},
  volume = {10},
  issue = {11},
  pages = {4637--4642},
  numpages = {0},
  year = {1974},
  month = {Dec},
  publisher = {American Physical Society},
  doi = {10.1103/PhysRevB.10.4637},
  url = {https://link.aps.org/doi/10.1103/PhysRevB.10.4637}
}

@article{PhysRevLett.70.3651,
  title = {Observation of the spin-Peierls transition in linear ${\mathrm{Cu}}^{2+}$ (spin-1/2) chains in an inorganic compound ${\mathrm{CuGeO}}_{3}$},
  author = {Hase, Masashi and Terasaki, Ichiro and Uchinokura, Kunimitsu},
  journal = {Phys. Rev. Lett.},
  volume = {70},
  issue = {23},
  pages = {3651--3654},
  numpages = {0},
  year = {1993},
  month = {Jun},
  publisher = {American Physical Society},
  doi = {10.1103/PhysRevLett.70.3651},
  url = {https://link.aps.org/doi/10.1103/PhysRevLett.70.3651}
}

@article{Kuboki_1987,
doi = {10.7567/JJAPS.26S3.593},
url = {https://dx.doi.org/10.7567/JJAPS.26S3.593},
year = {1987},
month = {jan},
publisher = {},
volume = {26},
number = {S3-1},
pages = {593},
author = {Kuboki, Kazuhiro and Fukuyama, Hidetoshi},
title = {Spin-Peierls Transition with Competing Interactions},
journal = {Japanese Journal of Applied Physics}
}

@article{PhysRevB.57.R14004,
  title = {Nonadiabatic approach to spin-Peierls transitions via flow equations},
  author = {Uhrig, G\"otz S.},
  journal = {Phys. Rev. B},
  volume = {57},
  issue = {22},
  pages = {R14004--R14007},
  numpages = {0},
  year = {1998},
  month = {Jun},
  publisher = {American Physical Society},
  doi = {10.1103/PhysRevB.57.R14004},
  url = {https://link.aps.org/doi/10.1103/PhysRevB.57.R14004}
}

@article{PhysRevB.60.6566,
  title = {Quantum lattice fluctuations in a frustrated Heisenberg spin-Peierls chain},
  author = {Wei\ss{}e, A. and Wellein, G. and Fehske, H.},
  journal = {Phys. Rev. B},
  volume = {60},
  issue = {9},
  pages = {6566--6573},
  numpages = {0},
  year = {1999},
  month = {Sep},
  publisher = {American Physical Society},
  doi = {10.1103/PhysRevB.60.6566},
  url = {https://link.aps.org/doi/10.1103/PhysRevB.60.6566}
}

@article{PhysRevLett.121.147201,
  title = {Quantization of the Thermal Hall Conductivity at Small Hall Angles},
  author = {Ye, Mengxing and Hal\'asz, G\'abor B. and Savary, Lucile and Balents, Leon},
  journal = {Phys. Rev. Lett.},
  volume = {121},
  issue = {14},
  pages = {147201},
  numpages = {5},
  year = {2018},
  month = {Oct},
  publisher = {American Physical Society},
  doi = {10.1103/PhysRevLett.121.147201},
  url = {https://link.aps.org/doi/10.1103/PhysRevLett.121.147201}
}

@article{PhysRevX.8.031032,
  title = {Approximately Quantized Thermal Hall Effect of Chiral Liquids Coupled to Phonons},
  author = {Vinkler-Aviv, Yuval and Rosch, Achim},
  journal = {Phys. Rev. X},
  volume = {8},
  issue = {3},
  pages = {031032},
  numpages = {13},
  year = {2018},
  month = {Aug},
  publisher = {American Physical Society},
  doi = {10.1103/PhysRevX.8.031032},
  url = {https://link.aps.org/doi/10.1103/PhysRevX.8.031032}
}

@article{PhysRevResearch.2.033180,
  title = {Phonon dynamics in the Kitaev spin liquid},
  author = {Ye, Mengxing and Fernandes, Rafael M. and Perkins, Natalia B.},
  journal = {Phys. Rev. Res.},
  volume = {2},
  issue = {3},
  pages = {033180},
  numpages = {14},
  year = {2020},
  month = {Aug},
  publisher = {American Physical Society},
  doi = {10.1103/PhysRevResearch.2.033180},
  url = {https://link.aps.org/doi/10.1103/PhysRevResearch.2.033180}
}

@article{PhysRevB.101.035103,
  title = {Phonon renormalization in the Kitaev quantum spin liquid},
  author = {Metavitsiadis, Alexandros and Brenig, Wolfram},
  journal = {Phys. Rev. B},
  volume = {101},
  issue = {3},
  pages = {035103},
  numpages = {7},
  year = {2020},
  month = {Jan},
  publisher = {American Physical Society},
  doi = {10.1103/PhysRevB.101.035103},
  url = {https://link.aps.org/doi/10.1103/PhysRevB.101.035103}
}

@misc{dantas2024phonondynamicssitedisorderedkitaev,
      title={Phonon dynamics in the site-disordered Kitaev spin liquid}, 
      author={Vitor Dantas and Wen-Han Kao and Natalia B. Perkins},
      eprint={2406.19140},
      archivePrefix={arXiv},
      url={https://arxiv.org/abs/2406.19140}, 
}

@article{ljlq-1kgl,
  title = {Spin-phonon coupling and thermal Hall effect in the Kitaev model},
  author = {Oh, Taekoo and Nagaosa, Naoto},
  journal = {Phys. Rev. B},
  volume = {112},
  issue = {8},
  pages = {L081104},
  numpages = {8},
  year = {2025},
  month = {Aug},
  publisher = {American Physical Society},
  doi = {10.1103/ljlq-1kgl},
  url = {https://link.aps.org/doi/10.1103/ljlq-1kgl}
}

@Article{Kasahara2018,
author={Kasahara, Y.
and Ohnishi, T.
and Mizukami, Y.
and Tanaka, O.
and Ma, Sixiao
and Sugii, K.
and Kurita, N.
and Tanaka, H.
and Nasu, J.
and Motome, Y.
and Shibauchi, T.
and Matsuda, Y.},
title={Majorana quantization and half-integer thermal quantum Hall effect in a Kitaev spin liquid},
journal={Nature},
year={2018},
month={Jul},
day={01},
volume={559},
number={7713},
pages={227-231},
issn={1476-4687},
doi={10.1038/s41586-018-0274-0},
url={https://doi.org/10.1038/s41586-018-0274-0}
}

@Article{Chen2024,
author={Chen, Lu
and Lefran{\c{c}}ois, {\'E}tienne
and Vallipuram, Ashvini
and Barth{\'e}lemy, Quentin
and Ataei, Amirreza
and Yao, Weiliang
and Li, Yuan
and Taillefer, Louis},
title={Planar thermal Hall effect from phonons in a Kitaev candidate material},
journal={Nature Communications},
year={2024},
month={Apr},
day={25},
volume={15},
number={1},
pages={3513},
issn={2041-1723},
doi={10.1038/s41467-024-47858-5},
url={https://doi.org/10.1038/s41467-024-47858-5}
}

@article{PhysRevB.110.125130,
  title = {Spin-Peierls instability of deconfined quantum critical points},
  author = {Hofmeier, David and Willsher, Josef and Seifert, Urban F. P. and Knolle, Johannes},
  journal = {Phys. Rev. B},
  volume = {110},
  issue = {12},
  pages = {125130},
  numpages = {13},
  year = {2024},
  month = {Sep},
  publisher = {American Physical Society},
  doi = {10.1103/PhysRevB.110.125130},
  url = {https://link.aps.org/doi/10.1103/PhysRevB.110.125130}
}

@Article{Seifert2024,
author={Seifert, Urban F. P.
and Willsher, Josef
and Drescher, Markus
and Pollmann, Frank
and Knolle, Johannes},
title={Spin-Peierls instability of the U(1) Dirac spin liquid},
journal={Nature Communications},
year={2024},
month={Aug},
day={19},
volume={15},
number={1},
pages={7110},
issn={2041-1723},
doi={10.1038/s41467-024-51367-w},
url={https://doi.org/10.1038/s41467-024-51367-w}
}

@misc{ferrari2024stabilityalgebraicspinliquids,
      title={Stability of algebraic spin liquids coupled to quantum phonons}, 
      author={Francesco Ferrari and Josef Willsher and Urban F. P. Seifert and Roser Valentí and Johannes Knolle},
      eprint={2410.16376},
      archivePrefix={arXiv},
      url={https://arxiv.org/abs/2410.16376}, 
}

@article{PhysRevB.104.035126,
  title = {Effects of spin-phonon coupling in frustrated Heisenberg models},
  author = {Ferrari, Francesco and Valent\'{\i}, Roser and Becca, Federico},
  journal = {Phys. Rev. B},
  volume = {104},
  issue = {3},
  pages = {035126},
  numpages = {9},
  year = {2021},
  month = {Jul},
  publisher = {American Physical Society},
  doi = {10.1103/PhysRevB.104.035126},
  url = {https://link.aps.org/doi/10.1103/PhysRevB.104.035126}
}

@article{PhysRevLett.89.037204,
  title = {Peierls-Like Transition Induced by Frustration in a Two-Dimensional Antiferromagnet},
  author = {Becca, Federico and Mila, Fr\'ed\'eric},
  journal = {Phys. Rev. Lett.},
  volume = {89},
  issue = {3},
  pages = {037204},
  numpages = {4},
  year = {2002},
  month = {Jul},
  publisher = {American Physical Society},
  doi = {10.1103/PhysRevLett.89.037204},
  url = {https://link.aps.org/doi/10.1103/PhysRevLett.89.037204}
}

@article{PhysRevB.104.165133,
  title = {Nesting instability of gapless U(1) spin liquids with spinon Fermi pockets in two dimensions},
  author = {Kr\"uger, Wilhelm G. F. and Janssen, Lukas},
  journal = {Phys. Rev. B},
  volume = {104},
  issue = {16},
  pages = {165133},
  numpages = {19},
  year = {2021},
  month = {Oct},
  publisher = {American Physical Society},
  doi = {10.1103/PhysRevB.104.165133},
  url = {https://link.aps.org/doi/10.1103/PhysRevB.104.165133}
}

@article{PhysRevLett.62.1694,
  title = {Valence-bond and spin-Peierls ground states of low-dimensional quantum antiferromagnets},
  author = {Read, N. and Sachdev, Subir},
  journal = {Phys. Rev. Lett.},
  volume = {62},
  issue = {14},
  pages = {1694--1697},
  numpages = {0},
  year = {1989},
  month = {Apr},
  publisher = {American Physical Society},
  doi = {10.1103/PhysRevLett.62.1694},
  url = {https://link.aps.org/doi/10.1103/PhysRevLett.62.1694}
}

@article{PhysRevB.42.4568,
  title = {Spin-Peierls, valence-bond solid, and N\'eel ground states of low-dimensional quantum antiferromagnets},
  author = {Read, N. and Sachdev, Subir},
  journal = {Phys. Rev. B},
  volume = {42},
  issue = {7},
  pages = {4568--4589},
  numpages = {0},
  year = {1990},
  month = {Sep},
  publisher = {American Physical Society},
  doi = {10.1103/PhysRevB.42.4568},
  url = {https://link.aps.org/doi/10.1103/PhysRevB.42.4568}
}

@misc{2024nonequilibrium,
      title={Tuning magnetic interactions with nonequilibrium optical phonon populations}, 
      author={Milan Kornjača and Rebecca Flint},
      eprint={2410.21373},
      archivePrefix={arXiv},
      url={https://arxiv.org/abs/2410.21373}, 
}

@article{PhysRevLett.115.177205,
  title = {Spin-Peierls Instability of Three-Dimensional Spin Liquids with Majorana Fermi Surfaces},
  author = {Hermanns, Maria and Trebst, Simon and Rosch, Achim},
  journal = {Phys. Rev. Lett.},
  volume = {115},
  issue = {17},
  pages = {177205},
  numpages = {6},
  year = {2015},
  month = {Oct},
  publisher = {American Physical Society},
  doi = {10.1103/PhysRevLett.115.177205},
  url = {https://link.aps.org/doi/10.1103/PhysRevLett.115.177205}
}

@article{PhysRevLett.133.223401,
  title = {Uncovering Emergent Spacetime Supersymmetry with Rydberg Atom Arrays},
  author = {Li, Chengshu and Liu, Shang and Wang, Hanteng and Zhang, Wenjun and Li, Zi-Xiang and Zhai, Hui and Gu, Yingfei},
  journal = {Phys. Rev. Lett.},
  volume = {133},
  issue = {22},
  pages = {223401},
  numpages = {8},
  year = {2024},
  month = {Nov},
  publisher = {American Physical Society},
  doi = {10.1103/PhysRevLett.133.223401},
  url = {https://link.aps.org/doi/10.1103/PhysRevLett.133.223401}
}

@article{PhysRevLett.126.206801,
  title = {Realization of Supersymmetry and Its Spontaneous Breaking in Quantum Hall Edges},
  author = {Ma, Ken K. W. and Wang, Ruojun and Yang, Kun},
  journal = {Phys. Rev. Lett.},
  volume = {126},
  issue = {20},
  pages = {206801},
  numpages = {7},
  year = {2021},
  month = {May},
  publisher = {American Physical Society},
  doi = {10.1103/PhysRevLett.126.206801},
  url = {https://link.aps.org/doi/10.1103/PhysRevLett.126.206801}
}

@article{PhysRevLett.117.166802,
  title = {All Majorana Models with Translation Symmetry are Supersymmetric},
  author = {Hsieh, Timothy H. and Hal\'asz, G\'abor B. and Grover, Tarun},
  journal = {Phys. Rev. Lett.},
  volume = {117},
  issue = {16},
  pages = {166802},
  numpages = {6},
  year = {2016},
  month = {Oct},
  publisher = {American Physical Society},
  doi = {10.1103/PhysRevLett.117.166802},
  url = {https://link.aps.org/doi/10.1103/PhysRevLett.117.166802}
}

@article{PhysRevLett.126.236802,
  title = {Boundary Supersymmetry of $(1+1)\mathrm{D}$ Fermionic Symmetry-Protected Topological Phases},
  author = {Prakash, Abhishodh and Wang, Juven},
  journal = {Phys. Rev. Lett.},
  volume = {126},
  issue = {23},
  pages = {236802},
  numpages = {6},
  year = {2021},
  month = {Jun},
  publisher = {American Physical Society},
  doi = {10.1103/PhysRevLett.126.236802},
  url = {https://link.aps.org/doi/10.1103/PhysRevLett.126.236802}
}

@article{PhysRevLett.90.120402,
  title = {Lattice Models with $\mathcal{N}=2$ Supersymmetry},
  author = {Fendley, Paul and Schoutens, Kareljan and de Boer, Jan},
  journal = {Phys. Rev. Lett.},
  volume = {90},
  issue = {12},
  pages = {120402},
  numpages = {4},
  year = {2003},
  month = {Mar},
  publisher = {American Physical Society},
  doi = {10.1103/PhysRevLett.90.120402},
  url = {https://link.aps.org/doi/10.1103/PhysRevLett.90.120402}
}

@article{PhysRevB.103.014435,
  title = {Lattice vibration as a knob on exotic quantum criticality},
  author = {Han, SangEun and Lee, Junhyun and Moon, Eun-Gook},
  journal = {Phys. Rev. B},
  volume = {103},
  issue = {1},
  pages = {014435},
  numpages = {9},
  year = {2021},
  month = {Jan},
  publisher = {American Physical Society},
  doi = {10.1103/PhysRevB.103.014435},
  url = {https://link.aps.org/doi/10.1103/PhysRevB.103.014435}
}

@article{PaulFendley_2003,
doi = {10.1088/0305-4470/36/50/004},
url = {https://dx.doi.org/10.1088/0305-4470/36/50/004},
year = {2003},
month = {dec},
publisher = {},
volume = {36},
number = {50},
pages = {12399},
author = {Paul Fendley and Bernard Nienhuis and Kareljan Schoutens},
title = {Lattice fermion models with supersymmetry},
journal = {Journal of Physics A: Mathematical and General}
}

@article{PhysRevLett.119.107202,
  title = {Edge Quantum Criticality and Emergent Supersymmetry in Topological Phases},
  author = {Li, Zi-Xiang and Jiang, Yi-Fan and Yao, Hong},
  journal = {Phys. Rev. Lett.},
  volume = {119},
  issue = {10},
  pages = {107202},
  numpages = {6},
  year = {2017},
  month = {Sep},
  publisher = {American Physical Society},
  doi = {10.1103/PhysRevLett.119.107202},
  url = {https://link.aps.org/doi/10.1103/PhysRevLett.119.107202}
}

@article{XiaoYang_2004,
doi = {10.1088/0305-4470/37/38/003},
url = {https://dx.doi.org/10.1088/0305-4470/37/38/003},
year = {2004},
month = {sep},
publisher = {},
volume = {37},
number = {38},
pages = {8937},
author = {Xiao Yang and Paul Fendley},
title = {Non-local spacetime supersymmetry on the lattice},
journal = {Journal of Physics A: Mathematical and General}
}

@article{PhysRevLett.95.046403,
  title = {Exact Results for Strongly Correlated Fermions in $2+1$ Dimensions},
  author = {Fendley, Paul and Schoutens, Kareljan},
  journal = {Phys. Rev. Lett.},
  volume = {95},
  issue = {4},
  pages = {046403},
  numpages = {4},
  year = {2005},
  month = {Jul},
  publisher = {American Physical Society},
  doi = {10.1103/PhysRevLett.95.046403},
  url = {https://link.aps.org/doi/10.1103/PhysRevLett.95.046403}
}

@article{PhysRevLett.101.146406,
  title = {Charge Frustration and Quantum Criticality for Strongly Correlated Fermions},
  author = {Huijse, Liza and Halverson, James and Fendley, Paul and Schoutens, Kareljan},
  journal = {Phys. Rev. Lett.},
  volume = {101},
  issue = {14},
  pages = {146406},
  numpages = {4},
  year = {2008},
  month = {Oct},
  publisher = {American Physical Society},
  doi = {10.1103/PhysRevLett.101.146406},
  url = {https://link.aps.org/doi/10.1103/PhysRevLett.101.146406}
}

@article{PhysRevB.84.115124,
  title = {Exact ground states of a staggered supersymmetric model for lattice fermions},
  author = {Huijse, L. and Moran, N. and Vala, J. and Schoutens, K.},
  journal = {Phys. Rev. B},
  volume = {84},
  issue = {11},
  pages = {115124},
  numpages = {15},
  year = {2011},
  month = {Sep},
  publisher = {American Physical Society},
  doi = {10.1103/PhysRevB.84.115124},
  url = {https://link.aps.org/doi/10.1103/PhysRevB.84.115124}
}

@article{PhysRevB.100.195146,
  title = {Supersymmetry in an interacting Majorana model on the kagome lattice},
  author = {Li, Chengshu and Lantagne-Hurtubise, \'Etienne and Franz, Marcel},
  journal = {Phys. Rev. B},
  volume = {100},
  issue = {19},
  pages = {195146},
  numpages = {10},
  year = {2019},
  month = {Nov},
  publisher = {American Physical Society},
  doi = {10.1103/PhysRevB.100.195146},
  url = {https://link.aps.org/doi/10.1103/PhysRevB.100.195146}
}

@article{PhysRevB.103.085130,
  title = {Unwinding fermionic symmetry-protected topological phases: Supersymmetry extension},
  author = {Prakash, Abhishodh and Wang, Juven},
  journal = {Phys. Rev. B},
  volume = {103},
  issue = {8},
  pages = {085130},
  numpages = {34},
  year = {2021},
  month = {Feb},
  publisher = {American Physical Society},
  doi = {10.1103/PhysRevB.103.085130},
  url = {https://link.aps.org/doi/10.1103/PhysRevB.103.085130}
}

@article{PhysRevB.110.165124,
  title = {Supersymmetry on the honeycomb lattice: Resonating charge stripes, superfrustration, and domain walls},
  author = {Wilhelm, Patrick H. and Kwan, Yves H. and L\"auchli, Andreas M. and Parameswaran, S. A.},
  journal = {Phys. Rev. B},
  volume = {110},
  issue = {16},
  pages = {165124},
  numpages = {14},
  year = {2024},
  month = {Oct},
  publisher = {American Physical Society},
  doi = {10.1103/PhysRevB.110.165124},
  url = {https://link.aps.org/doi/10.1103/PhysRevB.110.165124}
}

@Article{Cai2022,
author={Cai, M.-L.
and Wu, Y.-K.
and Mei, Q.-X.
and Zhao, W.-D.
and Jiang, Y.
and Yao, L.
and He, L.
and Zhou, Z.-C.
and Duan, L.-M.},
title={Observation of supersymmetry and its spontaneous breaking in a trapped ion quantum simulator},
journal={Nature Communications},
year={2022},
month={Jun},
day={14},
volume={13},
number={1},
pages={3412},
issn={2041-1723},
doi={10.1038/s41467-022-31058-0},
url={https://doi.org/10.1038/s41467-022-31058-0}
}

@article{6722-tf9c,
  title = {Supersymmetry dynamics on Rydberg atom arrays},
  author = {Liu, Shuo and Wu, Zhengzhi and Zhang, Shi-Xin and Yao, Hong},
  journal = {Phys. Rev. B},
  volume = {112},
  issue = {2},
  pages = {L020301},
  numpages = {7},
  year = {2025},
  month = {Jul},
  publisher = {American Physical Society},
  doi = {10.1103/6722-tf9c},
  url = {https://link.aps.org/doi/10.1103/6722-tf9c}
}

@Article{Li2017,
author={Li, Zi-Xiang
and Jiang, Yi-Fan
and Jian, Shao-Kai
and Yao, Hong},
title={Fermion-induced quantum critical points},
journal={Nature Communications},
year={2017},
month={Aug},
day={22},
volume={8},
number={1},
pages={314},
issn={2041-1723},
doi={10.1038/s41467-017-00167-6},
url={https://doi.org/10.1038/s41467-017-00167-6}
}

@article{PhysRevB.96.195162,
  title = {Fermion-induced quantum critical points in two-dimensional Dirac semimetals},
  author = {Jian, Shao-Kai and Yao, Hong},
  journal = {Phys. Rev. B},
  volume = {96},
  issue = {19},
  pages = {195162},
  numpages = {8},
  year = {2017},
  month = {Nov},
  publisher = {American Physical Society},
  doi = {10.1103/PhysRevB.96.195162},
  url = {https://link.aps.org/doi/10.1103/PhysRevB.96.195162}
}

@article{PhysRevB.101.085105,
  title = {Fermion-induced quantum critical point in Dirac semimetals: A sign-problem-free quantum Monte Carlo study},
  author = {Li, Bo-Hai and Li, Zi-Xiang and Yao, Hong},
  journal = {Phys. Rev. B},
  volume = {101},
  issue = {8},
  pages = {085105},
  numpages = {7},
  year = {2020},
  month = {Feb},
  publisher = {American Physical Society},
  doi = {10.1103/PhysRevB.101.085105},
  url = {https://link.aps.org/doi/10.1103/PhysRevB.101.085105}
}

@article{PhysRevB.67.245316,
  title = {Fermions, strings, and gauge fields in lattice spin models},
  author = {Levin, Michael and Wen, Xiao-Gang},
  journal = {Phys. Rev. B},
  volume = {67},
  issue = {24},
  pages = {245316},
  numpages = {10},
  year = {2003},
  month = {Jun},
  publisher = {American Physical Society},
  doi = {10.1103/PhysRevB.67.245316},
  url = {https://link.aps.org/doi/10.1103/PhysRevB.67.245316}
}

@article{PhysRevB.101.115113,
  title = {Microscopic definitions of anyon data},
  author = {Kawagoe, Kyle and Levin, Michael},
  journal = {Phys. Rev. B},
  volume = {101},
  issue = {11},
  pages = {115113},
  numpages = {24},
  year = {2020},
  month = {Mar},
  publisher = {American Physical Society},
  doi = {10.1103/PhysRevB.101.115113},
  url = {https://link.aps.org/doi/10.1103/PhysRevB.101.115113}
}

@article{PhysRevLett.73.2158,
  title = {Flux Phase of the Half-Filled Band},
  author = {Lieb, Elliott H.},
  journal = {Phys. Rev. Lett.},
  volume = {73},
  issue = {16},
  pages = {2158--2161},
  numpages = {0},
  year = {1994},
  month = {Oct},
  publisher = {American Physical Society},
  doi = {10.1103/PhysRevLett.73.2158},
  url = {https://link.aps.org/doi/10.1103/PhysRevLett.73.2158}
}

@article{PhysRevLett.107.066801,
  title = {Possible Lattice Distortions in the Hubbard Model for Graphene},
  author = {Frank, Rupert L. and Lieb, Elliott H.},
  journal = {Phys. Rev. Lett.},
  volume = {107},
  issue = {6},
  pages = {066801},
  numpages = {4},
  year = {2011},
  month = {Aug},
  publisher = {American Physical Society},
  doi = {10.1103/PhysRevLett.107.066801},
  url = {https://link.aps.org/doi/10.1103/PhysRevLett.107.066801}
}

@misc{seiberg1994powerholomorphyexact,
      title={The Power of Holomorphy -- Exact Results in 4D SUSY Field Theories}, 
      author={Nathan Seiberg},
      eprint={hep-th/9408013},
      archivePrefix={arXiv},
      url={https://arxiv.org/abs/hep-th/9408013}, 
}

@article{ANDERSON1973153,
title = {Resonating valence bonds: A new kind of insulator?},
journal = {Materials Research Bulletin},
volume = {8},
number = {2},
pages = {153-160},
year = {1973},
issn = {0025-5408},
doi = {https://doi.org/10.1016/0025-5408(73)90167-0},
url = {https://www.sciencedirect.com/science/article/pii/0025540873901670},
author = {P.W. Anderson}
}

@article{PhysRevLett.52.1575,
  title = {Conformal Invariance, Unitarity, and Critical Exponents in Two Dimensions},
  author = {Friedan, Daniel and Qiu, Zongan and Shenker, Stephen},
  journal = {Phys. Rev. Lett.},
  volume = {52},
  issue = {18},
  pages = {1575--1578},
  numpages = {0},
  year = {1984},
  month = {Apr},
  publisher = {American Physical Society},
  doi = {10.1103/PhysRevLett.52.1575},
  url = {https://link.aps.org/doi/10.1103/PhysRevLett.52.1575}
}

@article{PhysRevB.35.8865,
  title = {Topology of the resonating valence-bond state: Solitons and high-${T}_{c}$ superconductivity},
  author = {Kivelson, Steven A. and Rokhsar, Daniel S. and Sethna, James P.},
  journal = {Phys. Rev. B},
  volume = {35},
  issue = {16},
  pages = {8865--8868},
  numpages = {0},
  year = {1987},
  month = {Jun},
  publisher = {American Physical Society},
  doi = {10.1103/PhysRevB.35.8865},
  url = {https://link.aps.org/doi/10.1103/PhysRevB.35.8865}
}

@article{PhysRevLett.96.110405,
  title = {Detecting Topological Order in a Ground State Wave Function},
  author = {Levin, Michael and Wen, Xiao-Gang},
  journal = {Phys. Rev. Lett.},
  volume = {96},
  issue = {11},
  pages = {110405},
  numpages = {4},
  year = {2006},
  month = {Mar},
  publisher = {American Physical Society},
  doi = {10.1103/PhysRevLett.96.110405},
  url = {https://link.aps.org/doi/10.1103/PhysRevLett.96.110405}
}

@article{PhysRevLett.96.110404,
  title = {Topological Entanglement Entropy},
  author = {Kitaev, Alexei and Preskill, John},
  journal = {Phys. Rev. Lett.},
  volume = {96},
  issue = {11},
  pages = {110404},
  numpages = {4},
  year = {2006},
  month = {Mar},
  publisher = {American Physical Society},
  doi = {10.1103/PhysRevLett.96.110404},
  url = {https://link.aps.org/doi/10.1103/PhysRevLett.96.110404}
}

@article{PhysRevLett.83.195,
  title = {Spin-Peierls Transition in the Heisenberg Chain with Finite-Frequency Phonons},
  author = {Sandvik, Anders W. and Campbell, David K.},
  journal = {Phys. Rev. Lett.},
  volume = {83},
  issue = {1},
  pages = {195--198},
  numpages = {0},
  year = {1999},
  month = {Jul},
  publisher = {American Physical Society},
  doi = {10.1103/PhysRevLett.83.195},
  url = {https://link.aps.org/doi/10.1103/PhysRevLett.83.195}
}

@article{
doi:10.1126/science.1248253,
author = {Tarun Grover  and D. N. Sheng  and Ashvin Vishwanath },
title = {Emergent Space-Time Supersymmetry at the Boundary of a Topological Phase},
journal = {Science},
volume = {344},
number = {6181},
pages = {280-283},
year = {2014},
doi = {10.1126/science.1248253},
URL = {https://www.science.org/doi/abs/10.1126/science.1248253}}

@article{PhysRevLett.118.166802,
  title = {Emergence of Supersymmetric Quantum Electrodynamics},
  author = {Jian, Shao-Kai and Lin, Chien-Hung and Maciejko, Joseph and Yao, Hong},
  journal = {Phys. Rev. Lett.},
  volume = {118},
  issue = {16},
  pages = {166802},
  numpages = {6},
  year = {2017},
  month = {Apr},
  publisher = {American Physical Society},
  doi = {10.1103/PhysRevLett.118.166802},
  url = {https://link.aps.org/doi/10.1103/PhysRevLett.118.166802}
}

@article{PhysRevLett.114.237001,
  title = {Emergent Spacetime Supersymmetry in 3D Weyl Semimetals and 2D Dirac Semimetals},
  author = {Jian, Shao-Kai and Jiang, Yi-Fan and Yao, Hong},
  journal = {Phys. Rev. Lett.},
  volume = {114},
  issue = {23},
  pages = {237001},
  numpages = {5},
  year = {2015},
  month = {Jun},
  publisher = {American Physical Society},
  doi = {10.1103/PhysRevLett.114.237001},
  url = {https://link.aps.org/doi/10.1103/PhysRevLett.114.237001}
}

@article{
doi:10.1126/sciadv.aau1463,
author = {Zi-Xiang Li  and Abolhassan Vaezi  and Christian B. Mendl  and Hong Yao },
title = {Numerical observation of emergent spacetime supersymmetry at quantum criticality},
journal = {Science Advances},
volume = {4},
number = {11},
pages = {eaau1463},
year = {2018},
doi = {10.1126/sciadv.aau1463},
URL = {https://www.science.org/doi/abs/10.1126/sciadv.aau1463}}

@article{PhysRevLett.115.051601,
  title = {Bootstrapping the Three Dimensional Supersymmetric Ising Model},
  author = {Bobev, Nikolay and El-Showk, Sheer and Maz\'a\ifmmode \check{c}\else \v{c}\fi{}, Dalimil and Paulos, Miguel F.},
  journal = {Phys. Rev. Lett.},
  volume = {115},
  issue = {5},
  pages = {051601},
  numpages = {5},
  year = {2015},
  month = {Jul},
  publisher = {American Physical Society},
  doi = {10.1103/PhysRevLett.115.051601},
  url = {https://link.aps.org/doi/10.1103/PhysRevLett.115.051601}
}

@article{SupMat,
	Journal = {Please see the Supplementary Material at [URL] for the proof that the system always lies in the zero flux in the adiabatic limit, which includes the Refs. \cite{doi:10.1126/science.1248253,PhysRevB.76.075103,PhysRevLett.73.2158,PhysRevLett.107.066801,PhysRevB.90.245109,PhysRevLett.119.127204,PhysRevLett.115.051601,PhysRevLett.116.100402}. We also provide the details of our numerical calculations, the mean field analysis of the Kitaev-SSH model with a finite phonon frequency,  the RG analysis of the emergent SUSY, as well as the derivation of the temperature scaling of the thermal conductivity }
}

@article{GERVAIS1971632,
title = {Field theory interpretation of supergauges in dual models},
journal = {Nuclear Physics B},
volume = {34},
number = {2},
pages = {632-639},
year = {1971},
issn = {0550-3213},
doi = {https://doi.org/10.1016/0550-3213(71)90351-8},
url = {https://www.sciencedirect.com/science/article/pii/0550321371903518},
author = {J.-L. Gervais and B. Sakita}
}

@article{WESS197439,
title = {Supergauge transformations in four dimensions},
journal = {Nuclear Physics B},
volume = {70},
number = {1},
pages = {39-50},
year = {1974},
issn = {0550-3213},
doi = {https://doi.org/10.1016/0550-3213(74)90355-1},
url = {https://www.sciencedirect.com/science/article/pii/0550321374903551},
author = {J. Wess and B. Zumino}
}

@article{PhysRevB.106.024413,
  title = {Thermal Hall effect in the Kitaev-Heisenberg system with spin-phonon coupling},
  author = {Li, Shaozhi and Okamoto, Satoshi},
  journal = {Phys. Rev. B},
  volume = {106},
  issue = {2},
  pages = {024413},
  numpages = {7},
  year = {2022},
  month = {Jul},
  publisher = {American Physical Society},
  doi = {10.1103/PhysRevB.106.024413},
  url = {https://link.aps.org/doi/10.1103/PhysRevB.106.024413}
}

@article{PhysRevLett.102.217202,
  title = {Algebraic Spin Liquid in an Exactly Solvable Spin Model},
  author = {Yao, Hong and Zhang, Shou-Cheng and Kivelson, Steven A.},
  journal = {Phys. Rev. Lett.},
  volume = {102},
  issue = {21},
  pages = {217202},
  numpages = {4},
  year = {2009},
  month = {May},
  publisher = {American Physical Society},
  doi = {10.1103/PhysRevLett.102.217202},
  url = {https://link.aps.org/doi/10.1103/PhysRevLett.102.217202}
}

@article{DIMOPOULOS1981150,
title = {Softly broken supersymmetry and SU(5)},
journal = {Nuclear Physics B},
volume = {193},
number = {1},
pages = {150-162},
year = {1981},
issn = {0550-3213},
doi = {https://doi.org/10.1016/0550-3213(81)90522-8},
url = {https://www.sciencedirect.com/science/article/pii/0550321381905228},
author = {Savas Dimopoulos and Howard Georgi}
}

@article{PhysRevB.90.245109,
  title = {Conformal field theories at nonzero temperature: Operator product expansions, Monte Carlo, and holography},
  author = {Katz, Emanuel and Sachdev, Subir and S\o{}rensen, Erik S. and Witczak-Krempa, William},
  journal = {Phys. Rev. B},
  volume = {90},
  issue = {24},
  pages = {245109},
  numpages = {19},
  year = {2014},
  month = {Dec},
  publisher = {American Physical Society},
  doi = {10.1103/PhysRevB.90.245109},
  url = {https://link.aps.org/doi/10.1103/PhysRevB.90.245109}
}

@article{PhysRevLett.119.127204,
  title = {Thermal Transport in the Kitaev Model},
  author = {Nasu, Joji and Yoshitake, Junki and Motome, Yukitoshi},
  journal = {Phys. Rev. Lett.},
  volume = {119},
  issue = {12},
  pages = {127204},
  numpages = {6},
  year = {2017},
  month = {Sep},
  publisher = {American Physical Society},
  doi = {10.1103/PhysRevLett.119.127204},
  url = {https://link.aps.org/doi/10.1103/PhysRevLett.119.127204}
}

@article{PhysRevLett.116.100402,
  title = {Optical Conductivity of Topological Surface States with Emergent Supersymmetry},
  author = {Witczak-Krempa, William and Maciejko, Joseph},
  journal = {Phys. Rev. Lett.},
  volume = {116},
  issue = {10},
  pages = {100402},
  numpages = {5},
  year = {2016},
  month = {Mar},
  publisher = {American Physical Society},
  doi = {10.1103/PhysRevLett.116.100402},
  url = {https://link.aps.org/doi/10.1103/PhysRevLett.116.100402}
}

\clearpage
\newpage
\widetext

\title {Supplementary material for `Emergent spacetime supersymmetry at 2D fractionalized quantum criticality'}
\date{\today}

\maketitle

\onecolumngrid

\setcounter{equation}{0}
\setcounter{figure}{0}
\setcounter{table}{0}
\makeatletter
\renewcommand{\theequation}{S\arabic{equation}}
\renewcommand{\thefigure}{S\arabic{figure}}
\renewcommand{\bibnumfmt}[1]{[S#1]}
\renewcommand{\citenumfont}[1]{S#1}
\subsection{A. Lieb's theorems}
Here we apply two Lieb's theorems \cite{PhysRevLett.73.2158,PhysRevLett.107.066801} to prove that the ground state of the Kitaev-SSH model must lie in the zero flux sector of the phonon-mediated hoppings $t_{\langle ij\rangle}=\hat{u}_{\langle ij\rangle}(a_{\mu}+\hat{X}_{\langle ij\rangle})$ and  both the hopping module $\{t_{\langle ij\rangle}\}$ and phonon configuration $\{X_{ij}\}$ must respect the mirror symmetry of the model in the adiabatic limit $M\rightarrow +\infty$.

In the Majorana fermion representation, the Hamiltonian of the model in the adiabatic limit is:
\begin{equation}
\begin{aligned}
    \hat{H}=\sum_{\langle ij\rangle\in \mu}\hat{u}_{\langle ij\rangle}(a_{\mu}+X_{\langle ij\rangle})\left(i\hat{c}_i\hat{c}_j\right)+\sum_{\langle ij\rangle}\frac{1}{2\lambda}X^2_{\langle ij\rangle},
    \end{aligned}
    \label{adiabatic}
\end{equation}
where $a_x=a_y=1,a_z=a,0<a<1$. To technically facilitate Lieb's theorems, we consider a complex-fermion version of $\hat{H}$:
\begin{equation}
\begin{aligned}
    \hat{H}&= \hat{H}^f+\hat{H}^{\text{phonon}}\\
     &=\sum_{\langle ij\rangle\in \mu}t_{ij}\left(\hat{f}_{i}^{\dagger}\hat{f}_{j}+\text{h.c.}\right)+\sum_{\langle ij\rangle}\frac{1}{2\lambda}X^2_{\langle ij\rangle}.
\end{aligned}
\end{equation}
The ground state configurations of $t_{\langle ij\rangle}=\hat{u}_{\langle ij\rangle}(a_{\mu}+\hat{X}_{\langle ij\rangle})$ in these two models are the same since their ground state energies are the same $E_g(\{t_{ij}\})=E_f+\frac{1}{2\lambda}E_{\text{phonon}}$, where $E_f,E_{\text{phonon}}$ are the ground state energies of the fermion part $\sum_{\langle ij\rangle\in \mu}t_{\langle ij\rangle}\left(i\hat{c}_i\hat{c}_j\right)$ ( or $\sum_{\langle ij\rangle\in \mu}t_{\langle ij\rangle}\left(\hat{f}_{i}^{\dagger}\hat{f}_{j}+\text{h.c.}\right)$) and the phonon part $\sum_{\langle ij\rangle}X^2_{\langle ij\rangle}$,  respectively.

\begin{figure}
    \centering
    \includegraphics[width=0.5\linewidth]{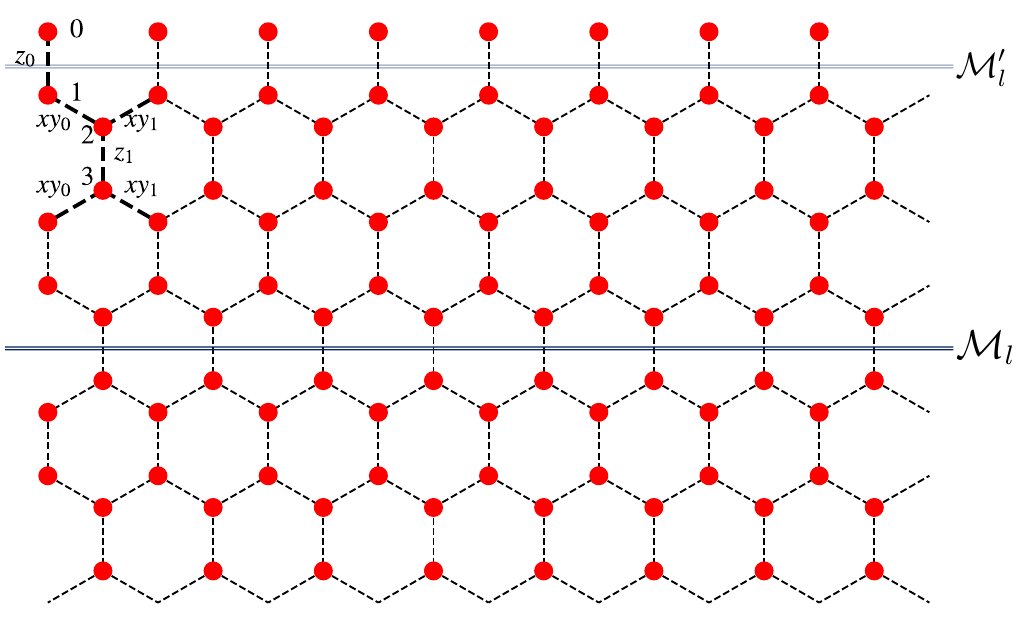}
    \caption{Schematic of the honeycomb lattice and its mirror symmetry. $\mathcal{M}_l$ denotes the mirror plane that maps the upper half of the system onto the lower half.
    According to Lieb’s theorem, the bonds along the vertical direction are equivalent, so the lattice retains translational symmetry in the vertical direction, but with an enlarged unit cell containing four sites. The six bonds in a unit cell are reduced to four independent degrees of freedom, as Lieb’s theorem constrains the values of the $xy$-bonds.
    When periodic boundary conditions are applied in the vertical direction, the mirror plane appears in pairs ($\mathcal M_l, \mathcal M_l'$) and also divide the system into two parts that are mapped onto each other by the mirror symmetry.
    }
    \label{fig:S1}
\end{figure}
We note that the Hamiltonian $\hat{H}$ respects the mirror symmetry with mirror planes $\mathcal{M}_l$ bisecting any row-$l$ of $z$-type of bonds (as shown in Fig.~\ref{fig:S1}), and we first take all the $t_{ij}$ on the $z$-bonds in the $l$-th row to be positive, which is always possible due to the $\mathbb{Z}_2$ gauge transformations of $\hat{u}_{ij}$. We divide the $\hat{H}$ into three parts: $\hat{H}=\hat{H}_{\text{lower}}+\hat{H}_{\text{upper}}+\hat{H}_{\text{int}}$, where $\hat{H}_{\text{lower}}(\hat{H}_{\text{upper}})$ only contains the terms with all the lattice sites in the lower (upper) half of the mirror plane $\mathcal{M}_l$. $\hat{H}_{\text{int}}$ contains the terms on the bonds bisected by the mirror plane $\mathcal{M}_l$.

Next, we define a transformation $\mathcal{R}_l$ for each of the mirror plane $\mathcal{M}_l$ following \cite{PhysRevLett.73.2158}. $\mathcal{R}_l$ is the composition of two transformations: (1) Unitary particle-hole transformation: $\hat{f}_i\rightarrow\hat{f}_i^{\dagger}$; (2) Mirror transformation across the mirror plane $\mathcal{M}_l$. Let us begin with the purely fermion part $\hat{H}^f$, then we have the following inequality according to Lieb's theorem \cite{PhysRevLett.73.2158}:
\begin{equation}
    (\tr[e^{-\beta(\hat{H}^f_{\text{lower}}+\hat{H}^f_{\text{upper}}+\hat{H}^f_{\text{int}})}])^2\leq \left(\tr[e^{-\beta(\hat{H}^f_{\text{lower}}+\mathcal{R}_l[\hat{H}^f_{\text{lower}}]+\hat{H}^f_{\text{int}})}]\right)\left(\tr[e^{-\beta(\hat{H}^f_{\text{upper}}+\mathcal{R}_l[\hat{H}^f_{\text{upper}}]+\hat{H}^f_{\text{int}})}]\right),
\end{equation}
Further, we can find that symmetric phonon configuration provides an upper bound of the total partition function $Z(\{u_{ij},X_{ij}\})=e^{-\sum_{ij}\frac{\beta}{2\lambda}X^2_{ij}}\tr[e^{-\beta(\hat{H}^f_{\text{lower}}+\hat{H}^f_{\text{upper}}+\hat{H}^f_{\text{int}})}]$ , or equivalently the lower bound of the free energy $f=-\frac{\ln(Z(\{u_{ij},X_{ij}\}))}{\beta}$ : 
\begin{equation}
    \begin{aligned}
        (e^{-\sum_{ij}\frac{\beta}{2\lambda}X^2_{ij}}\tr[e^{-\beta(\hat{H}^f_{\text{lower}}+\hat{H}^f_{\text{upper}}+\hat{H}^f_{\text{int}})}]&)^2\leq\left(\tr[e^{-\beta(\hat{H}^f_{\text{lower}}+\mathcal{R}_l[\hat{H}^f_{\text{lower}}]+\hat{H}^f_{\text{int}})}]e^{-\beta \left[E_{\text{phonon}}(\{X_{ij\in\text{lower}},\mathcal{R}_l[X_{ij\in\text{lower}}],X_{ij\in\text{int}}\})\right]}\right) \\
        &\cdot\left(\tr[e^{-\beta(\hat{H}^f_{\text{upper}}+\mathcal{R}_l[\hat{H}^f_{\text{upper}}]+\hat{H}^f_{\text{int}})}]e^{-\beta \left[E_{\text{phonon}}(\{\mathcal{R}_l[X_{ij\in\text{upper}}],X_{ij\in\text{upper}},X_{ij\in\text{int}}\})\right]}\right),
    \end{aligned}
\end{equation}
where we use the identity $\frac{2}{2\lambda}\sum_{\langle ij\rangle}X^2_{\langle ij\rangle}=\frac{1}{2\lambda}\sum_{\langle ij\rangle\in \text{lower}}X^2_{\langle ij\rangle}+\frac{1}{2\lambda}\sum_{\langle ij\rangle\in \text{lower}}\mathcal{R}_l[X_{\langle ij\rangle}]^2+\left(\text{lower}\leftrightarrow \text{upper}\right)+\frac{2}{2\lambda}\sum_{\langle ij\rangle\in \text{int}}X^2_{\langle ij\rangle}$.
We note that in the zero temperature limit, this gives the lower bound of the ground state energy $E_g=-\lim_{T\rightarrow0}\frac{\ln(Z)}{\beta}$ : $E_g(\hat{H}_{\text{lower}}+\hat{H}_{\text{upper}}+\hat{H}_{\text{int}})\geq\frac{1}{2}\left[ E_g(\mathcal{R}_l[\hat{H}_{\text{upper}}]+\hat{H}_{\text{upper}}+\hat{H}_{\text{int}})+E_g(\hat{H}_{\text{lower}}+\mathcal{R}_l[\hat{H}_{\text{lower}}]+\hat{H}_{\text{int}})\right]$. As a result, the optimal energy is achieved by the zero flux sector of $\{t_{ij}\}$ and the symmetric phonon configuration $\{X_{ij}\}$ with respect to any mirror plane. 
Given that the flux of $\{t_{ij}\}$ is zero, all the $t_{ij}$ can be made positive through $\mathbb{Z}_2$ gauge transformations, and thus they are invariant under any mirror $\mathcal{M}_l$ transformations since $\{X_{ij}\}$ are symmetric.


\subsection{B. Numerical simulation details}
\begin{figure}
    \centering
    \includegraphics[width=1.0\linewidth]{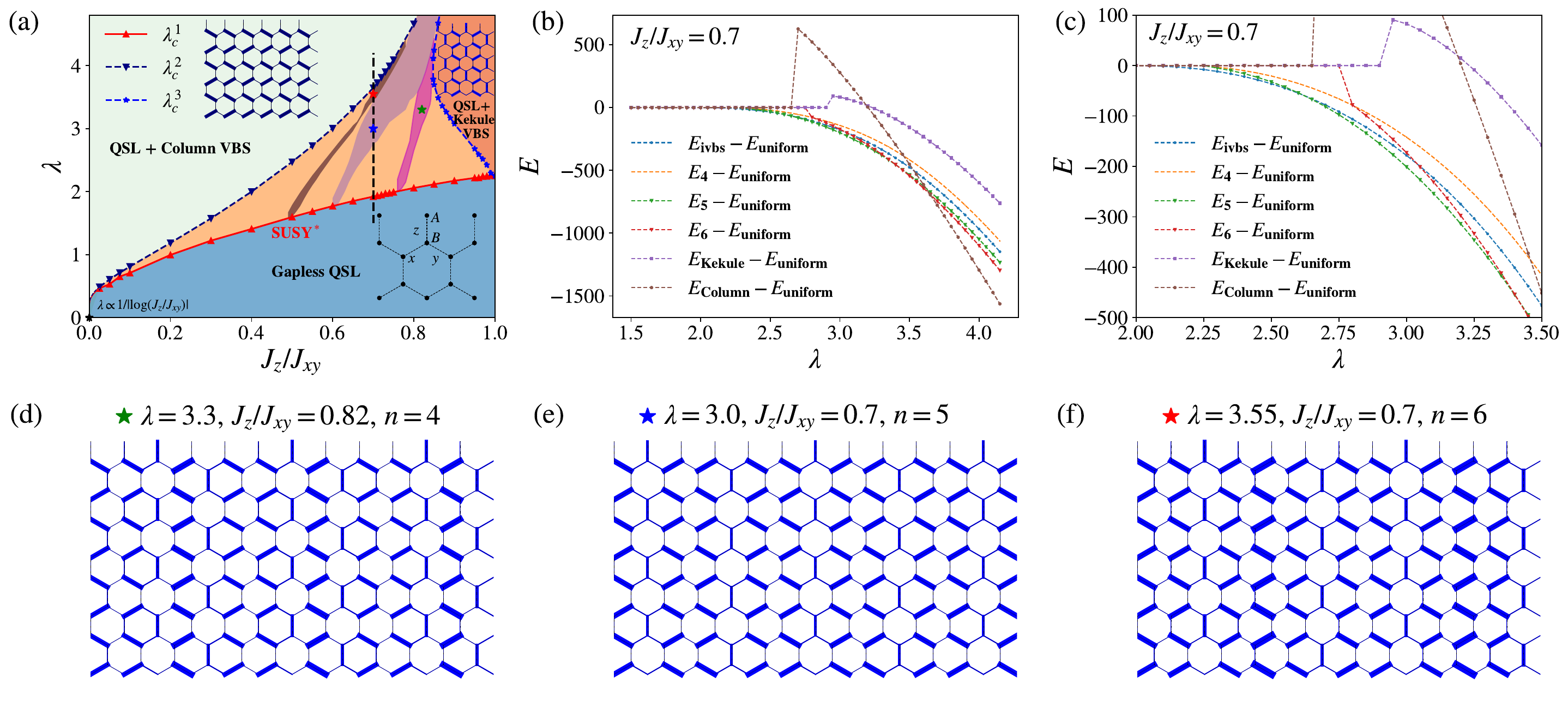}
    \caption{(a) Detailed phase diagram including commensurate VBS phases with small periodicities, indicated by the shaded regions.
    (b) Comparison of the energies obtained from different self-consistent setups at $J_z/J_{xy}=0.7$, corresponding to the vertical dashed line in (a). The simulations are performed on a $2 \times 240 \times 120$ lattice. All energies are shown relative to the uniform-constraint energy. 
    (c) Zoomed-in view of panel (b).
    (d–f) Phonon-field configurations at three representative points in the commensurate VBS phases with $n=4,5,6$, corresponding to the star markers in (a). The critical points from the gapless spin liquid to the commensurate VBS-spin liquids with periods $n=5,6$ have emergent SUSY.
    }
    \label{fig:S2}
\end{figure}

In this section, we present details of the numerical simulations used to obtain the phase diagram shown in Fig.~1. The ground-state configuration is determined by minimizing the energy of the Hamiltonian in Eq. \eqref{adiabatic}, which leads to the self-consistent equation $X_{ij} = -\lambda\left<f_i^\dagger f_j+\text{h.c.}\right>$. 
As proven in the previous section, Lieb’s theorem guarantees the reflection symmetry along the mirror planes shown in Fig.~\ref{fig:S1}. Consequently, the number of independent phonon degrees of freedom $X_{\braket{ij}}$ is greatly reduced. In our numerical calculations, we employ the lattice geometry shown in Fig. \ref{fig:S1}, with periodic boundary conditions along both the vertical and horizontal directions. Lieb’s theorem ensures that the configuration respects translational symmetry along the vertical direction. Consequently, for a system of size $2\times 2L_y\times L_x$, the number of free parameters is reduced from $6 L_x L_y$ to only $4L_x$: there are only four inequivalent bonds along the y-(vertical) direction, while they remain independent along x. We  denote these four bonds as $z_0,z_1,xy_0,xy_1$ in  Fig. \ref{fig:S1}.
In the simulation, we iteratively apply the self-consistent equations to the phonon fields until the energy converges, i.e., when the energy difference between two successive iterations is smaller than a prescribed tolerance (we use $10^{-7}$ in calculations).
Nevertheless, as with most mean-field–like methods, the converged configuration may correspond to a local minimum rather than the global minimum, especially in the presence of first-order transitions. To mitigate this issue, we adopt several strategies to optimize the configuration.
First, we restrict the phonon fields to be spatially uniform and denote the corresponding converged energy by $E_\text{uniform}$. 
To detect possible incommensurate VBS order, we then relax all restrictions on the phonon fields and start from weak coupling $\lambda$. As the coupling is varied step by step, we initialize each new simulation using the configuration converged at the previous coupling, supplemented by a small random perturbation (specifically, each field is multiplied by a random factor drawn from $[0.9,1.1]$). We denote the resulting converged energy by $E_\text{ivbs}$.
Because the perturbations are small, this procedure may fail to detect commensurate VBS order as the coupling increases. To address this issue, we explicitly restrict the phonon fields to adopt the Kekul\'e (column) pattern shown in Fig.~\ref{fig:S2}(a), and denote the corresponding converged energy by $E_\text{kekule}$ ($E_\text{column}$). In addition, to probe VBS order with small commensurate periodicities, we impose a fixed periodicity on the phonon fields and denote the resulting energy by $E_n$ where $n$ is the imposed period along the horizontal direction. The Kekule (column) pattern is actually equivalent to $n=3$ (1).

In Fig.~\ref{fig:S2}(b,c), we present the energies obtained from the different approaches at $\lambda = 0.7$ (corresponding to the vertical dashed line in Fig.~\ref{fig:S2}(a)), taking $E_{\text{uniform}}$ as the reference.
At weak coupling, all energies coincide with $E_{\text{uniform}}$, indicating the absence of VBS order. 
As $\lambda$ increases, $E_{\text{ivbs}}$ becomes lower than the uniform energy, signaling the onset of incommensurate VBS order. 
Upon further increasing $\lambda$, VBS states with small commensurate periodicities become energetically favorable relative to the incommensurate state, and eventually the columnar VBS state emerges as the ground state.
In Fig.~\ref{fig:S2}(a), the regions corresponding to commensurate VBS phases with $n=4,5,6$ are highlighted by shaded areas. 
In Fig.~\ref{fig:S2}(d–f), we also show the phonon-field patterns at three representative points within these commensurate phases.
Although the precise phase boundaries are limited by the finite lattice resolution, their extent along the transition line to the gapless spin liquid is expected to shrink to zero in the thermodynamic limit.
Moreover, from the renormalization-group analysis, the critical points to spin liquids with  commensurate VBS orders have emergent SUSY when the VBS periods are larger than 4, which means the critical points to the commensurate VBS-spin liquids with periods  $n=5,6$ have emergent SUSY.

\begin{figure}
    \centering
    \includegraphics[width=0.45\linewidth]{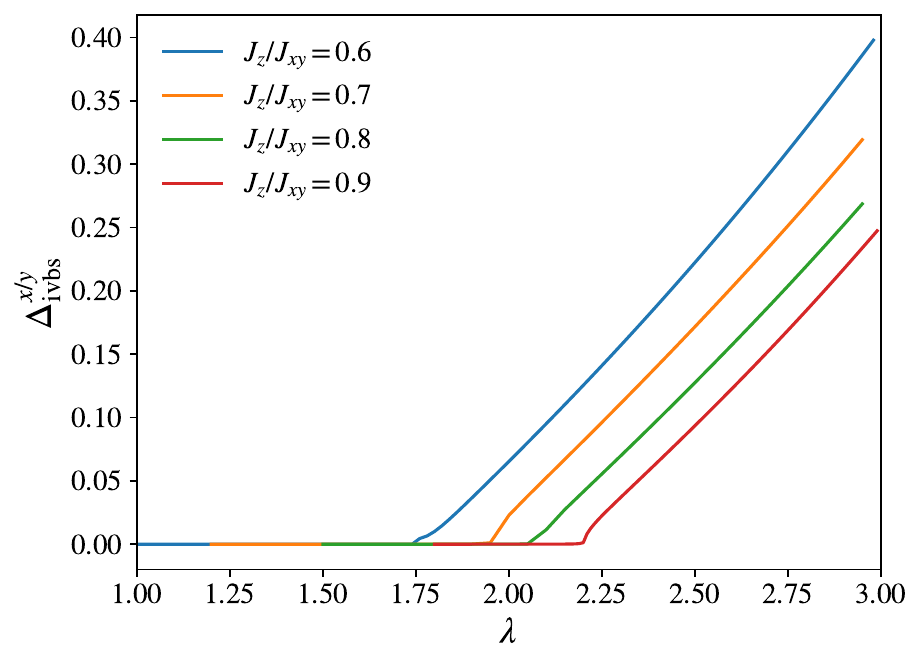}
    \caption{The incommensurate VBS order parameter defined on the $x/y$-bonds is given by $\Delta^{x/y}_\text{ivbs}\equiv \big|\tfrac{1}{N}\sum_{i\in A} e^{-2i  \tilde {\mathbf{K}} \cdot\mathbf r_i} X_{\langle ij\rangle\in x/y}\big|$. The simulations are performed on a $2\times 240\times 120$ lattice. The resulting behavior is qualitatively identical to that of the $z$-bond order parameter shown in Fig.~1(b).}
    \label{fig:S3}
\end{figure}

The phase boundary of the columnar VBS phase is determined by the crossing point between $E_\text{column}$ and the lowest energy among the other competing states. 
In contrast, the phase boundary of the incommensurate VBS phase cannot be identified from an energy crossing, as the transition is expected to be continuous. 
Instead, we determine this phase boundary by examining the behavior of the corresponding order parameter $\Delta_\text{ivbs}$, as shown in Fig.~1(b). 
Note that the definition of $\Delta_\text{ivbs}$ given in the caption of Fig.~1 is not unique to the $z$-bond. 
One may equivalently define order parameters using the $x/y$-bond, $\Delta^{x/y}_\text{ivbs}\equiv \big|\tfrac{1}{N}\sum_{i\in A} e^{-2i  \tilde {\mathbf{K}} \cdot\mathbf r_i} X_{\langle ij\rangle\in x/y}\big|$. 
Since all of these order parameters detect the same translational-symmetry breaking, they simultaneously appear at the same critical coupling and exhibit qualitatively identical behavior, as shown in Fig.~\ref{fig:S3}.
Here $\tilde{\mathbf{K}}$ denotes the momentum of the Dirac point, which can shift with the coupling $\lambda$. In practice, we therefore use the nonzero-momentum Fourier component with the largest amplitude. 
We have verified that this momentum is consistent with the Dirac-point momentum at the transition point.

\subsection{C. Perturbative analyses for finite phonon frequency $\omega_{\text{ph}}$}
Here, we consider the Kitaev-SSH model with small but finite phonon frequency $\omega_{\text{ph}}=\sqrt{\frac{k}{m}}>0$, and perform a perturbative analysis to show 
the robustness of the adiabatic limit phase diagram and our proposed schematic phase diagram with finite $\omega_{\text{ph}}$.
First, the $\mathbb{Z}_2$ gauge fields are always exactly deconfined and static in the Kitaev-SSH model even with a finite $\omega_{\text{ph}}$, so the ground state is always a $\mathbb{Z}_2$ spin liquid without any magnetic order, since the spin operator $\tau^{\mu}_i=i\hat{c}_i^{\mu}\hat{c}_i$ always flips two conserved $\mathbb{Z}_2$ fluxes and thus $\langle\tau_i^{\mu} \rangle=0$. 
Further, the $\mathbb{Z}_2$ flux in the adiabatic limit  always has a gap of order $J$, where $J$ is the bare Kitaev interaction. 
As a result, we consider the limit $0<\omega_{\text{ph}}\ll J$, and the ground state still lies in the zero flux sector. Therefore, we only need to consider the itinerant Majorana fermions $\hat{c}_i$ coupled with phonons, which we expect the leading instability can only be various VBS orders due to the fact that $\hat{c}_i$ does not possess other internal degrees of freedom that could give rise to competing instabilities. 
Hence, we first use the following mean-field ansatz:
\begin{equation}
    \hat{H}_{\text{MF}}=\sum_{\langle ij\rangle} J_{\langle ij\rangle} (i\hat{c}_i\hat{c}_j)+\frac{\hat{P}_{\langle ij\rangle}^2}{2m}+\frac{k}{2} (\hat{X}_{\langle ij\rangle}-\bar{X}_{\langle ij\rangle})^2,
\end{equation}
whose ground state $| \psi(J_{\braket{ij}},\bar X)\rangle$ is the variational mean field wave function for the following interacting Kitaev-SSH model:
\begin{equation}
 \hat{H}_f=\sum_{\langle ij\rangle\in \mu}\left[(J_{\mu}+g\hat{X}_{\langle ij\rangle})\right]\left(i\hat{c}_i\hat{c}_j\right)+\sum_{\langle ij\rangle}\frac{\hat{P}^2_{\langle ij\rangle}}{2m}+\frac{k}{2}\hat{X}^2_{\langle ij\rangle}.
 \label{lattice}
\end{equation}
This ansatz is sufficiently general because the bond-dependent variational parameters $J_{\langle ij\rangle},\bar{X}_{\langle ij\rangle}$ capture all VBS patterns considered above. These parameters are determined by minimizing the mean-field energy $\braket{\hat H_f}_\text{MF}$
which evaluates to 
\begin{equation}
    E_\text{MF}=\sum_{\braket{ij}\in\mu}(J_\mu+g\braket{\hat X_{\braket{ij}}}_\text{MF})\braket{i \hat c_i\hat c_j}_\text{MF}+\sum_{\langle ij\rangle}\frac{k}{2}\bar{X}^2_{\langle ij\rangle}  +\frac {1}{2}N_b \hbar \omega_{\text{ph}},
\end{equation}
where $N_b$ is the number of bonds on the honeycomb lattice. The mean field expectation value of $\braket{\hat X_{\braket{ij}}}_\text{MF}=\bar X_{\braket{ij}}$. 
With fixed $\bar X_{\langle ij\rangle}$, minimizing the fermionic part of the Hamiltonian $\sum_{\braket{ij}\in\mu}(J_\mu+g{\bar X_{\braket{ij}}})i \hat c_i\hat c_j$, requires that $J_{\braket{ij}}=J_\mu+g{\bar X_{\braket{ij}}}$.
This leads to the self-consistent mean-field equation for $\bar X$:
\begin{equation}
\bar X_{ij} = -g/k\left<i\hat c_i \hat c_j\right>_{J_{\braket{ij}}=J_\mu + g\bar X_{ij}},
\end{equation}
which is similar 
to that obtained in the adiabatic limit (but with a rescaling of $X$).
The finite phonon frequency $\omega_{\text{ph}}$ therefore contributes 
an additive constant $\frac{1}{2} N_b\hbar \omega_{\text{ph}}$ to the mean-field energy, and does not explicitly modify the self-consistent solution. Consequently, at the mean-field level, no additional lower-energy phases emerge at finite $\omega_{\text{ph}}$, and the adiabatic-limit phase diagram remains robust in the regime $\omega_{\text{ph}}\ll J$.

At first glance, it may 
seem 
that the critical spin-phonon coupling $g_c$ is independent of the phonon frequency $\omega_{\text{ph}}$. Rather, this indicates that our mean-field treatment correctly captures the leading contribution with  a finite $\omega_{\text{ph}}$. Formally, we can perturbatively expand the critical value in terms of the phonon frequency $\omega_\text{ph}$,  $g_c(\omega_{\text{ph}})=\sum_n \omega_{\text{ph}}^n g_{c,n}$, and our above calculation has captured the leading term $g_{c,0}$. 
The mean-field calculation can be refined in a controlled way to incorporate the higher-order corrections in $\omega_{\text{ph}}$.

To qualitatively capture the effect of a  finite phonon frequency $\omega_{\text{ph}}$ on the phase boundary, i.e. whether the critical $g_c$ increases or decreases with the increase of the frequency $\omega_{\text{ph}}$,  we adopt a continuum approximation to the lattice model in Eq. \eqref{lattice}. Here we only focus on the phase boundary between the gapless QSL and the gapped QSL with the incommensurate VBS order, since this is the critical line where our predicted emergent SUSY appears. We will show that the phase boundary predicted by our mean field theory is consistent with that in Fig. 1(c) in the main text, where a larger $g_c$ is needed with the increase of $\omega_{\text{ph}}$.

The path integral and the action in the continuum limit takes:
\begin{eqnarray}
Z&=&\int D\psi D\bar{\psi}D\phi D\phi^* e^{-S},\quad S = S_f + S_b + S_{\rm int} \,,\\
    S_f &=& \int d \tau d^{2} x{\psi}^\dagger( \partial_{\tau}-iv_{x}\partial_x\sigma_y-iv_{y}\partial_y\sigma_x )\psi \,,\\
    S_b &=& \int d \omega d^2x \phi^*(\omega,x) \left(M\omega^2+K \right) \phi(\omega,x)\,,\\
    S_{\text{int}}&=&g\int d\tau d^2x\left(\phi \psi^{T} \sigma_y \psi+\text{h.c.}\right)\,,
\end{eqnarray}
where $\psi,\phi$ are the continuum limit of the lattice Majorana fermions $\hat{c}$ and the phonon mode near momentum $-2\tilde{\mathbf{K}}$ respectively, as we derived in the main text. We now treat the bosonic field $\phi(\omega,\vec{k}=0)$ as the order parameter and make the mean field approximation. As will become clear below, this approximation is controlled by the small parameter $\omega_{\text{ph}}$ and becomes exact in the adiabatic limit $\omega_{\text{ph}}\rightarrow 0$. To ensure that our mean-field treatment is smoothly connected to the exact adiabatic result, we first recall that in the adiabatic limit $\omega_{\text{ph}}=0$ the bosonic field $\phi(\omega,\vec{k})$ reduces to $\phi(\omega,\vec{x})=\phi_0\delta(\omega)$ in the adiabatic limit $\omega_{\text{ph}}=0$, since the kinetic energy $\int d\tau d^2x M|\partial_{\tau}\phi(
\tau,\vec{x})|^2$ only allow a time-independent $\phi(\tau,\vec{x})=\phi(\vec{x})$ configuration in the adiabatic limit $M\rightarrow\infty$; equivalently, in frequency space, the field $\phi(\omega,\vec{x})=\phi_0(\vec{x})\delta(\omega)$ contains a delta function $\delta(\omega)$. Moreover, in the ground-state sector $\phi_0(\vec{x})=\phi_0$ is spatially uniform: a constant configuration gaps the Dirac cone and minimizes the energy, in agreement with the lattice numerical results.  For a finite $\omega_{\text{ph}}$, the $\delta(\omega)$ is broadened into a smooth function peaked at $\omega=0$ with a characteristic width set by $2\omega_{\text{ph}}$. This claim can be justified by considering the effective interaction between fermions if we integrate out the phonon modes:
\begin{equation}
    V_{\text{eff}}=-\frac{g^2}{K}\int d\Omega d\omega d\omega^{\prime}d^2x\left( \frac{\omega_{\text{ph}}^2}{\Omega^2+\omega_{\text{ph}}^2}\right) (\bar{\psi}(\omega,x)\sigma_y\bar{\psi}^T(-\omega+\Omega,x)) (\psi^T(\omega^{\prime},x)\sigma_y\psi(-\omega^{\prime}+\Omega,x)),
\end{equation}
where the interaction form factor $\frac{\omega_{\text{ph}}^2}{\Omega^2+\omega_{\text{ph}}^2}\approx \Theta(\omega_{\text{ph}}-|\Omega|)$ is a smooth function centered at $\omega=0$ with width $2\omega_{\text{ph}}$ when $\omega_{\text{ph}}$ is small. Physically, this means that the fermion bilinear $\int d\omega d^2x\psi^T(\omega^{\prime},x)\sigma_y\psi(-\omega^{\prime}+\Omega,x)$, which can also be viewed as the order parameter similar to the phonon field $\phi (\vec{k}=0,\Omega)$, can only have nonzero expectation values in the interval $|\Omega|\leq \omega_{\text{ph}}$, since the fermions have attractive interactions only in this regime. As a result, both $\phi (\vec{k}=0,\Omega)$ and $\int d\omega d^2x\psi^T(\omega^{\prime},x)\sigma_y\psi(-\omega^{\prime}+\Omega,x)$ can be treated as proportional to $\Theta(\omega_{\text{ph}}-|\Omega|)$.

Hence, we approximate the field $\phi(\omega,\vec{x})$ as:
\begin{equation}
    \phi(\omega,\vec{x})\approx \phi_0\left[\frac{1}{2\omega_{\text{ph}}}(\Theta(\omega_{\text{ph}}-|\omega|))\right],
\end{equation}
where $\phi_0$ is a constant which is viewed as the order parameter, and the coefficient $\frac{1}{2\omega_{\text{ph}}}$ is set to make the function $\lim_{\omega_{\text{ph}}\rightarrow 0}\left[\frac{1}{2\omega_{\text{ph}}}(\Theta(\omega_{\text{ph}}-|\omega|))\right]=\delta(\omega)$. Now we can integrate out the fermion modes, and keep the effective action (free energy) to the second order of $\phi_0$:
\begin{equation}
\begin{aligned}
    Z_{\text{MF}}&=e^{-V\frac{f}{\omega_{\text{ph}}}}\\
f&=K|\phi_0|^2\int^{\infty}_{-\infty} d\omega\frac{\Theta(\omega_{\text{ph}}-|\omega|)}{2\omega_{\text{ph}}}\left[\frac{\omega^2/\omega^2_{\text{ph}}+1}{2}-\frac{g^2}{K}\chi_{\text{pp}}(0,\omega)\right],
\end{aligned}
\end{equation}
where $V$ is the total area of the system and the particle-particle susceptibility $\chi_{\text{pp}}(0,\omega)$ is:
\begin{equation}
\begin{aligned}
        \chi_{\text{pp}}(0,\omega)=\int d^2k d\Omega\,\tr\left[\sigma_y G(\vec{k},\Omega)\sigma_yG(-\vec{k},-\Omega+\omega)\right],
\end{aligned}
\end{equation}
where $G(\vec{k},\Omega)$ is the fermion Green's function. We  can see that the mean-field approximation is controlled by the small parameter $\omega_{\text{ph}}$. In the path-integral formulation, the mean-field partition function takes the form $Z_{\text{MF}}=e^{-V\frac{f}{\omega_{\text{ph}}}}$,  so the free energy is effectively weighted by a prefactor $1/\omega_{\text{ph}}$. For small 
$\omega_{\text{ph}}$, this large prefactor suppresses fluctuations around the saddle point, making the saddle-point contribution dominant. In the adiabatic limit $\omega_{\text{ph}}\rightarrow 0$, the prefactor $1/\omega_{\text{ph}}\rightarrow \infty$, and fluctuations are completely suppressed; consequently, the mean-field solution becomes exact.

We  can directly obtain the coefficient of $|\phi_0|^2$ as:
\begin{eqnarray}
    f=|\phi_0|^2K\left(\frac{2}{3}-4\pi^2\lambda(1-\frac{\omega_{\text{ph}}\pi}{8\Lambda  }+O(\omega_{\text{ph}}^2))\right),
\end{eqnarray}
where $\Lambda$ is the momentum cutoff of the continuum model $|\vec{k}|\leq\Lambda$, and $\lambda=\frac{g^2\Lambda}{K}$ is the dimensionless spin-phonon coupling. 
For simplicity, we have set the unit $v_x=v_y=1$. 
Now the critical spin-phonon coupling is determined by the point where the coefficient of the second order term in the free energy $f$ changes sign:
\begin{equation}
   6\pi^2 \lambda_c=\frac{1}{1-\frac{\omega_{\text{ph}}\pi}{8\Lambda}+O(\omega^2_{\text{ph}})}\,,
\end{equation}
which is consistent with the phase boundary between the gapless QSL and the gapped QSL with the incommensurate VBS order shown in Fig. 1 (c) in the maintext.
\subsection{D. RG analysis for the emergent supersymmetry}

In this section, we provide the detail of the RG analysis for the emergent supersymmetry (SUSY) at the QCP. 
Our method closely follows that of Ref.~\cite{doi:10.1126/science.1248253,PhysRevB.76.075103}.
The effective field theory at the QCP is given by 
\begin{eqnarray}
    S &=& S_f + S_b + S_{\rm int} \,, \\
    S_f &=& \int d \tau d^{2} x{\psi}^\dagger( \partial_{\tau}-iv_{x}\partial_x\sigma_y-iv_{y}\partial_y\sigma_x )\psi \,,\\
    S_b &=& \int d \tau d^{2} x \left(\left|\partial_{\tau} \phi\right|^{2}+\sum_{i=x,y}v_{b,i}^{2}|\partial_i \phi|^{2}+u |\phi|^4 \right) \,,\\
    S_{\text{int}}&=&g\int d\tau d^2x\left(\phi \psi^{T} \sigma_y \psi+\text{h.c.}\right)\,,
\end{eqnarray}
where $\psi$ and $\phi$ are the Dirac fermion and the complex boson, respectively. 
$v_{i}$ and $v_{b,i}$ denote the velocities of the Dirac fermion and the complex boson, respectively. 

Before we dive into the detailed RG analyses, we briefly compare our model with those of \cite{doi:10.1126/science.1248253,PhysRevB.76.075103} to clarify the underlying physics.  The main difference is that we allow for velocity anisotropy: $v_x\neq v_y, v_{b_x}\neq v_{b_y}$, while Ref. \cite{doi:10.1126/science.1248253,PhysRevB.76.075103} work in the isotropic limit: $v_x= v_y=v, v_{b_x}= v_{b_y}=v_b$. Importantly, this anisotropy does not destroy the SUSY fixed point: our RG analysis below shows that the velocities flow toward: $v_x=v_{b,x},v_y=v_{b,y}$.  At that fixed point, SUSY can be made explicit by an anisotropic rescaling of coordinates: $x_i \rightarrow x^{\prime}_i=v_ix_i$. Another difference is that Ref. \cite{doi:10.1126/science.1248253,PhysRevB.76.075103} study, respectively, a single Majorana cone and two Dirac cones; however, the velocity renormalization is insensitive to the number of Dirac cones, so this distinction is not essential for our discussion.

%

We first show that the velocities flow to the fixed point  $v_x=v_{b,x},v_y=v_{b,y}$, which means the kinetic energy part flows to the SUSY fixed point. Following the methods in \cite{doi:10.1126/science.1248253,PhysRevB.76.075103}, we use the $\epsilon$-expansion momentum shell Wilson RG and work in the spacetime dimension $D=4-\epsilon$. 
The one-loop RG equations for the velocities are:
\begin{equation}
    \frac{dv_{i}}{dl}=2g^2N_Dv_{i}\left[\int \frac{\sin\theta d\theta d\phi}{4\pi}\left(\frac{f^2b-v^2_{i}\hat{p}^2_i(2f+b)}{bf^3(f+b)^2}\right)\right]\,,\quad \frac{dv_{b,i}}{dl}=-\frac{2g^2N_D}{v_{b,i}}\int \frac{\sin\theta d\theta d\phi}{4\pi}\left(\frac{v^2_{b,i}}{8f^3}-\frac{v^2_{i}}{2f^3}+\frac{9v_i^4\hat{p}_i^2}{8f^5}\right) \,.
    \label{velocity_rg}
\end{equation}
where $i=x,y$, $\hat{p}_x=\sin\theta\cos\phi,\hat{p}_y=\sin\theta\sin\phi,\hat{p}_z=\cos\theta$ and $f=\sqrt{\sum_jv^2_{j}\hat{p}^2_j}$, $b=\sqrt{\sum_jv^2_{b,j}\hat{p}^2_j}$. The parameter $N_D=\frac{A(S^{D-1})}{(2\pi)^D}$, which should be analytically continued to the physical spacetime dimension $D=3$, where $A(S^{D-1})$ is the area of the sphere $S^{D-1}$.

As a sanity check, we can take the isotropic limit $v_x= v_y=v$, $ v_{b_x}= v_{b_y}=v_b$ in our RG equations in~\eqref{velocity_rg}. 
In this limit, our RG equations reduce to the standard isotropic form and agree with Ref. \cite{doi:10.1126/science.1248253,PhysRevB.76.075103}:
\begin{equation}
    \frac{dv}{dl}=\frac{4g^2N_D(v_b-v)}{3v_b(v+v_b)^2},\quad \frac{dv_{b}}{dl}=-\frac{g^2N_D(v^2_b-v^2)}{4v_bv^3}.
    \label{isotropic_rg}
\end{equation}
The velocity RG equations in Ref. \cite{doi:10.1126/science.1248253} only differ from our results in \eqref{isotropic_rg} by a total factor of 4, since they consider a Majorana cone and real boson there; while the RG equations in Ref. \cite{PhysRevB.76.075103} are the same as ours by setting $D=3$ and  hence $N_D=\frac{1}{2\pi^2}$ here.

Now we return to our RG equations with anisotropic velocities  in Eq. \eqref{velocity_rg} and show that there is a stable fixed point $v_x=v_{b,x},v_y=v_{b,y}$. First, it is straightforward to verify that $v_i=v_{b,i}$ is indeed a fixed point (or fixed plane more exactly) of the RG equations Eq. \eqref{velocity_rg}, and we can further verify that this fixed point is a stable fixed point. We define the velocity difference variables $\Delta v_i=v_i-v_{b,i}$ and linearize the RG equations in  Eq. \eqref{velocity_rg} around the fixed point $\Delta v_i=0$ : 
\begin{equation}
    \frac{d \delta(\Delta v_i)}{dl}=-\frac{5g^2N_D}{6|v_xv_y|}\delta(\Delta v_i),
    \label{linear_rg}
\end{equation}
where $\delta(\Delta v_i)$ is the variation of $\Delta v_i$ around the fixed point $\Delta v_i=0$.  From the linearized RG equation in \eqref{linear_rg}, it is obvious that the fixed point $v_i = v_{b,i}$ is indeed a  stable fixed point, since any deviation $\delta(\Delta v_i)$ from the fixed point $v_i = v_{b,i}$ flows to zero, as long  as the Yukawa coupling $g\neq0$ at the fixed point. 

As we stated above, physically, the fixed point means that the kinetic energy is supersymmetric at the fixed point $v_i=v_{b,i}$ .  
First, the velocity anisotropy can be removed by a nonsingular anisotropic rescaling of coordinates $x_i \rightarrow x_i' = (v_i/v) x_i$.
In these coordinates, the kinetic terms become isotropic with velocity $v$, i.e. $v_i \partial_{x_i} \rightarrow v \partial_{x_i'}$. 
Finally, by an overall rescaling of space, we choose units with $v=1$, and equivalently, we have arrived at $v_x = v_y = 1$. 
This choice only changes nonuniversal metric factors and therefore does not affect the universality class. 
In these units the IR fixed-point action is rotationally invariant in spacetime, exhibiting emergent Lorentz symmetry---equivalently, the kinetic energy is supersymmetric.

Having established emergent Lorentz invariance/supersymmetry in the kinetic terms at the fixed point, we are now ready to study the RG flow of the interaction couplings. We will show that the couplings flow to nonzero values and, more importantly, that the total action has the emergent spacetime SUSY. At this stage, there is no intrinsic difference between the analysis here and those in Ref. \cite{doi:10.1126/science.1248253,PhysRevB.76.075103}.
Using the $4-\epsilon$ method, the one-loop RG equations for the coupling constants $g$ and $u$ in  are 
\begin{eqnarray}
    \frac{{\rm d} g^2}{{\rm d}l} &=& \epsilon g^2 - \frac32 g^4\,, \\
    \frac{{\rm d} u}{{\rm d}l} &=& \epsilon u +2 g^4 - \frac52 u^2 - g^2 u \,,
\end{eqnarray}
where $g^2 \rightarrow (g^2/2\pi^2)$, $ u \rightarrow (u/2\pi^2), $ were assumed implicitly. 
The stable fixed point of these RG equations is $u^\ast = g^{\ast2} = \frac23 \epsilon$. 
The nonzero values of the coupling constants characterize a strongly interacting fixed point, and also justify the assumption in the analysis of the velocity renormalization. 

We show that the total action at the fixed point is supersymmetric. 
After using the freedom to rescale spatial coordinates so that the fermion and boson velocities are isotropic, $v_x=v_y=v_{b,x}=v_{b,y}=1$, the action at the fixed point with $u=g^2$ can be written as $S=\int d\tau\,d^2x\,\mathcal L$ with
\begin{equation}
\mathcal L = |\partial_\tau\phi|^2+|\partial_x\phi|^2+|\partial_y\phi|^2 +\bar \psi\gamma^\mu\partial_\mu\psi +g\big(\phi\,\psi^T\sigma_y\psi+\phi^*\,\psi^\dagger\sigma_y(\psi^\dagger)^T\big)
+ g^2 |\phi|^4\,,
\end{equation}
where we have introduced the gamma matrices $\gamma^\tau=\sigma_z$, $\gamma^x=-\sigma_x$, $ \gamma^y=\sigma_y$, and $\bar \psi = \psi^\dag \gamma^\tau$. 
This theory is the $2+1$d $\mathcal N=2$ SUSY Wess--Zumino model. 
It is straightforward to show that the action is invariant under the SUSY transformations
\begin{eqnarray}
&& \delta\phi=\bar \varepsilon \psi\,, \qquad  \delta\psi=-(\gamma^\mu\partial_\mu\phi)\varepsilon+g(\phi^*)^2\varepsilon\,, 
\end{eqnarray}
where $\varepsilon$ is a Grassman number and plays the role of the SUSY transformation parameter. 
Therefore, we have shown that the QCP has emergent spacetime SUSY. 

\subsection{E. Temperature scaling of the thermal conductivity}
Here we investigate the temperature scaling of the thermal conductivity at the SUSY critical point following the logic in \cite{PhysRevB.90.245109}. The finite frequency thermal conductivity can be obtained from the imaginary frequency Kubo formula: $\kappa_{\text {Kubo }}^{\mu\nu}(\omega_n)=-\frac{1}{\omega_n TV}\int_0^{\beta}\langle T_{\tau}\left(J_Q^{\mu}(\tau)J_Q^{\nu}(0)\right)\rangle e^{-i\omega_n\tau} d\tau$ through the analytical continuation $\omega_n\rightarrow -i\omega+\delta$, where $\mu,\nu$ represents the spatial directions and the time-dependent total heat current is: $J_{Q}^{\mu}(\tau)=e^{\hat{H} \tau} J_{Q}^{\mu} e^{-\hat{H} \tau}$, while $\hat{H}$ is the Hamiltonian. $T,V$ are the temperature and volume of the system respectively. We first prove that the real part of our $\kappa^{\mu\nu}(\omega_n)$  is the same as that defined in the literature: $\Re[\kappa^{\mu \nu}(\omega)]=(1 / T V) \Re[\int_{0}^{\infty} d t e^{i(\omega+i \delta) t} \int_{0}^{\beta} d \lambda\left\langle J_{Q}^{\mu}(\lambda) J_{Q}^{\nu}(it)\right\rangle]$ \cite{PhysRevLett.119.127204}. This can be proved by using the Lehmann (spectral) representation. Firstly, $\Re[\kappa^{\mu\nu}(\omega)]$ is:
\begin{equation}
    \begin{aligned}
      \Re[\kappa^{\mu\nu}(\omega)]&=  \frac{1}{ZTV}\Re\left[\sum_{m,n}e^{-\beta E_n}\int_0^{\beta}d\lambda\int_0^{+\infty}dt e^{i(\omega+i\delta)t} \langle  n|J_Q^{\mu}(\lambda)|m\rangle\langle m|J_Q^{\nu}(it)|n\rangle\right]\\
      &= \frac{1}{ZTV}\Re\left[\sum_{m,n}e^{-\beta E_n}\langle  n|J_Q^{\mu}|m\rangle\langle m|J_Q^{\nu}|n\rangle\int_0^{\beta}d\lambda\int_0^{+\infty}dt e^{i(\omega+i\delta)t}e^{\lambda(E_n-E_m)} e^{i(E_m-E_n)t}\right]\\
      &= \frac{1}{ZTV}\Re\left[\sum_{m,n}\langle  n|J_Q^{\mu}|m\rangle\langle m|J_Q^{\nu}|n\rangle\frac{e^{-\beta E_m}-e^{-\beta E_n}}{E_m-E_n}\frac{1}{i\omega+i(E_m-E_n)-\delta}\right]\\
      &=\frac{1}{ZTV}\sum_{m,n}\langle  n|J_Q^{\mu}|m\rangle\langle m|J_Q^{\nu}|n\rangle\frac{e^{-\beta E_m}-e^{-\beta E_n}}{E_m-E_n}\left(-\pi\delta(\omega+E_m-E_n)\right).
    \end{aligned}
\end{equation}

Secondly, the Kubo formula is:
\begin{equation}
    \begin{aligned}
    \kappa_{\text {Kubo }}^{\mu\nu}(\omega_n)&=-\frac{1}{\omega_n TV}\int_0^{\beta}e^{-i\omega_n\tau}\langle  T_{\tau}\left(J_Q^{\mu}(\tau)J_Q^{\nu}(0)\right)\rangle d\tau\\
        &=-\frac{1}{\omega_nTV}\frac{1}{Z}\sum_{mn}e^{-\beta E_n}\langle n| J_Q^{\mu}|m\rangle\langle m|J_Q^{\nu}|n\rangle\int_0^{\beta}d\tau e^{-i\omega_n\tau} e^{\tau(E_n-E_m)}\\
        &=-\frac{1}{\omega_nTV}\frac{1}{Z}\sum_{mn}\langle n| J_Q^{\mu}|m\rangle\langle m|J_Q^{\nu}|n\rangle\frac{e^{-\beta E_m}-e^{-\beta E_n}}{-i\omega_n+E_n-E_m}.\\
    \end{aligned}
\end{equation}
After the analytical continuation, it becomes:
\begin{equation}
    \begin{aligned}
       \kappa_{\text {Kubo }}^{\mu\nu}(\omega) &=-\frac{1}{ZTV}\sum_{mn}\langle n| J_Q^{\mu}|m\rangle\langle m|J_Q^{\nu}|n\rangle\frac{e^{-\beta E_m}-e^{-\beta E_n}}{(-i\omega+\delta)(-\omega+E_n-E_m-i\delta)}\\
        &=-\frac{1}{ZTV} \sum_{mn}\langle n| J_Q^{\mu}|m\rangle\langle m|J_Q^{\nu}|n\rangle\left[\frac{e^{-\beta E_m}-e^{-\beta E_n}}{E_m-E_n}\pi\delta(\omega+E_m-E_n)+...\right],
        \label{imaginary}
    \end{aligned}
\end{equation}
when $...$ represents the imaginary part and $\omega\neq 0$. Now it is clear that the real parts of our $\kappa_{\text {Kubo }}^{\mu\nu}(\omega)$ and $\Re[\kappa^{\mu\nu}(\omega)]$ defined in \cite{PhysRevLett.119.127204} are the same for a nonzero frequency.

Following the logic in \cite{PhysRevB.90.245109}, the temperature scaling of the thermal conductivity in the high frequency limit $\frac{\omega}{T}\gg 1$ can be obtained through the OPE of heat current operator, so we first complete the time integral in $-\frac{1}{\omega_n TV}\int_0^{\beta}\langle T_{\tau}\left(J_Q^{i}(\tau)J_Q^{i}(0)\right)\rangle d\tau$:
\begin{equation}
    \begin{aligned}
       & -\frac{1}{\omega_n TV}\int_0^{\beta}e^{-i\omega_n\tau}\langle  T_{\tau}\left(J_Q^{\mu}(\tau)J_Q^{\nu}(0)\right)\rangle d\tau\\
        &=-\frac{\beta}{\omega_nV}\frac{1}{\beta^2}\sum_{\omega_1,\omega_2}\int_0^{\beta}e^{i\omega_1\tau}\langle J_Q^{i}(\omega_1,0)J_Q^{i}(\omega_2,0)\rangle e^{-i\omega_n\tau} d\tau\\
        &=-\frac{1}{\omega_nV\beta}\sum_{\omega_1,\omega_2}\langle J_Q^{i}(\omega_1,0)J_Q^{i}(\omega_2,0)\rangle \delta_{\omega_1-\omega_n,0}\beta\\
        &=-\frac{1}{\omega_nV}\langle J_Q^{i}(\omega_n,0)J_Q^{i}(-\omega_n,0)\rangle,
    \end{aligned}
\end{equation}
where $J_Q^{i}(\tau)=\frac{1}{\beta }\sum_{\omega_n}e^{i\omega_n\tau}J_Q^{i}(\omega_n,\vec{p}=0)$.

Now we return to our supersymmetric critical point. Since Majorana fermions and phonon do not have chemical potential, the heat current is the energy current: $J^i_Q(\tau)=\int d^2x T^{i0}(x,\tau)$, where $T^{i0}(x,\tau)$ is the energy-momentum tensor. The zero temperature OPE of $J^i_Q(\omega)$ takes:
\begin{equation}
    \begin{aligned}
        \lim_{|\mathbf{\omega}|\gg|\mathbf{p}|} J_Q^{i}(\mathbf{\omega},0)J_Q^i(-\mathbf{\omega}+\mathbf{p})=-\delta^3(p)|\omega|^3\kappa_0+\frac{c}{\omega^{\Delta-3}}|\phi|^2(\vec{p})+...,
    \end{aligned}
\end{equation}
where $\kappa_0=0$ is the zero temperature thermal conductivity: $-\delta^3(p)|\omega|^3\kappa_0=\delta^3(p)\left[\int d^2xd\tau e^{-i\omega\tau}\langle T^{i0}(x,\tau)T^{i0}(0)\rangle\right]=0$, and $\Delta=3-\frac{1}{\nu}\approx 1.9098$ is the scaling dimension of the bosonic field $|\phi|^2(\vec{x},\tau)$ at the critical point \cite{PhysRevLett.115.051601,PhysRevLett.116.100402}. We have neglected more irrelevant terms contributed  by the energy-momentum tensor with scaling dimension $\Delta_T=3$, since we are interested in the leading temperature scaling. As a result, the thermal conductivity scaling is: $\kappa^{ii}(\omega)\propto (i\omega)^{2-\Delta} T^{\Delta}$ after the analytical continuation of the frequency.

\end{document}